\documentclass[pra,twocolumn]{revtex4}\usepackage{psfrag}
\usepackage{epsfig}
\usepackage{amsmath}
\usepackage{amssymb}
\usepackage[colorlinks=true,linkcolor=blue,citecolor=blue]{hyperref}

\newcommand{\bra}[1]{\ensuremath{\langle#1|}}
\newcommand{\ket}[1]{\ensuremath{|#1\rangle}}

\newcommand{\be}{\begin{equation}}
\newcommand{\ee}{\end{equation}}
\newcommand{\avg}[1]{\ensuremath{\langle #1 \rangle}}

\newcommand{\im}{\text{i}}

\newcommand{\ie}{{\it i.e.}}

\newcommand{\eg}{{\it e.g. }}

\newcommand{\eqcite}[1]{Eq.~\eqref{#1}}

\newcommand{\nn}{\bold{n}}

\newcommand{\rr}{\bold{r}}
\newcommand{\RR}{\bold{R}}
\newcommand{\qq}{\bold{q}}

\newcommand{\BB}{\bold{B}}
\newcommand{\mm}{\bold{m}}

\newcommand{\KK}{\bold{K}}

\newcommand{\FF}{\bold{F}}

\newcommand{\kk}{\bold{k}}

\newcommand{\LAT}{\text{lat}}

\begin{document}

\title{Superconducting Vortex Lattices for Ultracold Atoms}

\author{O. Romero-Isart$^{1}$}
\author{C. Navau$^{2}$}
\author{A. Sanchez$^{2}$}
\author{P. Zoller$^{3,4}$}
\author{J. I. Cirac$^1$}

\affiliation{$^1$Max-Planck-Institut f\"ur Quantenoptik,
Hans-Kopfermann-Strasse 1,
D-85748, Garching, Germany.}
\affiliation{$^2$Grup d'Electromagnetisme, Departament de F\'isica, Universitat Aut\`onoma de Barcelona, 08193 Bellaterra, Barcelona, Catalonia, Spain. }
\affiliation{$^{3}$Institute for Theoretical Physics, University of Innsbruck, A-6020 Innsbruck, Austria.}
\affiliation{$^{4}$Institute for Quantum Optics and Quantum Information of the
Austrian Academy of Sciences, A-6020 Innsbruck, Austria.}

\begin{abstract}

We propose and analyze a nanoengineered vortex array in a thin-film type-II superconductor as a magnetic lattice for ultracold atoms. This proposal addresses several of the key questions in the development of atomic quantum simulators. By trapping atoms close to the surface, tools of nanofabrication and structuring of lattices on the scale of few tens of nanometers become available with a corresponding benefit in energy scales and temperature requirements. This can be combined with the possibility of magnetic single site addressing and manipulation together with a favorable scaling of superconducting surface-induced decoherence.

\end{abstract}

\maketitle

The ability to trap and manipulate ultracold atoms in lattice structures has led to remarkable experimental progress to build quantum simulators for Hubbard models, which are paradigmatic  in condensed-matter physics. A prominent example is atoms in optical lattices (OLs)~\cite{Bloch08a,Jaksch98}.  
When loading an ultracold  gas of neutral atoms into a lattice potential, atoms are positioned at the local minima of the lattice potential. In this situation, atoms can tunnel to neighboring lattice sites with tunnel coupling $t$ and interact on site due to short-ranged collisional interactions with a strength $U$~\cite{Jaksch98}~
\footnote{For bosonic atoms, the many-body dynamics are described by the Hubbard Hamiltonian
$\hat H= - t \sum_{\avg{ij}} ( \hat a^{\dagger}_{i}  \hat a_{j} +  \hat a^{\dagger}_{j}  \hat a_{i} ) +U\sum_{i} \hat n_{i} (\hat n_{i}-1)/2$,
where $\hat a_{i} (\hat a^{\dagger}_{i})$ annihilates (creates) a localized bosonic atom on site $i$, and $\hat n_{i}=\hat a^{\dagger}_{i} \hat a_{i}$.  The first term describes tunneling between neighboring sites with tunnel coupling $t$, and the second on-site interactions with strength $U$.  }.
 One of the significant interests for studying these type of models, in particular for spin=1/2 fermionic atoms~\cite{Hofstetter02,Greif12}, lies in the fact that in the strong coupling regime $U/t \gg 1$, superexchange processes (with coupling strengths $\sim t^{2}/U$) provide the basic mechanism for an antiferromagnetic coupling between spins on neighboring sites, which is closely related to studies of high-$T_\text{c}$ superconductivity within the Hubbard model~\cite{Lee06}.

An important challenge to simulate Bose- and Fermi-Hubbard Hamiltonians in a regime not accessible to classical computers~\cite{Cirac12} is the development of better cooling schemes in order to reduce the entropy of the simulator~\cite{Bloch12}. In particular, one demands $k_\text{B} T, \hbar \Gamma \ll t^{2}/U \ll t < E_\text{R}$, where $T$ is the temperature of the system ($k_\text{B}$ the Boltzmann's constant), $\Gamma$ the decoherence rate of the atoms, and $E_\text{R}=h^{2}/(8 m_\text{a} a^{2} )$ the recoil energy, where $m_\text{a}$ is the mass of the trapped atoms, $a$ the interlattice site distance, and $h=2 \pi \hbar $ the Planck's constant. In OLs, $a$ corresponds to half the optical wavelength, which leads to $E_\text{R}/k_\text{B} \sim 10^{-7}$~K (\eg for rubidium and wavelength $852$~nm). With present cooling techniques atoms can be prepared at few nanokelvins, which render the above set of inequalities very tight. Alternatively to designing better cooling schemes, one could loosen the above set of inequalities by reducing $a$ and thereby boosting the energy scale of the physical parameters of the quantum simulator. Because of the diffraction limit, this requires to trap atoms near a surface without adding new sources of decoherence~\cite{Gullans12}.

Here, we propose and analyze a new approach to trap and manipulate ultracold neutral atoms in arbitrary (periodic and nonperiodic) lattice potentials based on using a magnetic nanolattice  generated by a controlled array of superconducting vortices in thin-film type-II superconductors~\cite{Blatter94}.  
With present technologies, superconducting vortices can be positioned in complex structures by artificially pinning them in nanoengineered arrays of, for instance, completely etched holes (antidots) of various sizes and shapes; see Ref.~\cite{Wordenweber10} and references therein. 
Our proposal hints at the possibility to exploit this technology to fabricate and structure arbitrary  magnetic lattices for atomic physics at the fundamental length scales associated with superconducting vortices, the coherence length $\xi$,  and London's penetration depth $\lambda$, which can be of a few tens of nanometers. Moreover, the combination of all-magnetic trapping and manipulation and superconducting surfaces leads, in principle, to very favorable scalings on surface-induced decoherence, as discussed below. These features make this proposal significantly distinctive from previous schemes for magnetic lattices~\cite {Grabowski03,Singh08,Whitlock09,Abdelrahman10,Leung11}, where denser lattices with low decoherence are challenging, as well as for magnetic traps where an atom is trapped by the field created by macroscopic currents flowing in type-II superconducting materials~\cite{Shimizu09, Mueller10,Mueller10b, Zhang12,Siercke12,Markowsky12} and not by the field created by few controlled superconducting vortices, as proposed here.

We consider a superconducting film of thickness $d  \lesssim \lambda$.  The film is a type-II superconductor, \ie,~$\lambda/\xi >1/\sqrt{2}$,  where the value of $\xi$ typically  ranges between few to tens of nanometers~\cite{DegennesBook, Blatter94}. In thin films, the effective penetration depth is given by $\Lambda = \lambda^{2}/d \gtrsim \lambda \gtrsim d> \xi$, which hence can potentially be as small as few tens of nanometers. The upper side of the film is situated at the $x-y$ plane with $z=0$ and  contains a nanoengineered array of artificial pinning centers consisting of antidots of radius $R \gtrsim \xi$~\cite{Wordenweber10,Blatter94} distributed in a Bravais lattice $\RR=n_{1} \bold{a}_{1} + n_{2} \bold{a}_{2}$, where $\bold{a}_{1(2)}$ are the lattice primitive vectors and $n_{1(2)}$ range through all integer values. The density of antidots is $1/a^{2}$, where $a^{2}=|\bold{a}_{1} \times \bold{a}_{2}|$ is the area of the primitive cell. By cooling the film in the presence of an external field whose flux density is commensurate to the density of antidots, a single vortex is pinned in each antidot (pinning multiple vortices in each antidot is also possible, see \cite{Berdiyorov06}). This assumes that the density of antidots does not exceed $1/a^{2} \lesssim 1/\Lambda^{2}$, otherwise the vortex-vortex repulsion would prevent the vortices from sitting at the antidot lattice; see Supplemental Material (SM) for further discussion. Once the film is cooled and the vortex lattice is prepared, the external magnetic field is switched off and the vortices remain, each of them with a magnetic flux given by $\Phi_{0}=h/(2e)$, where $e$ is the charge of the electron.

\begin{figure*}
\begin{center}
\includegraphics[width=  \linewidth]{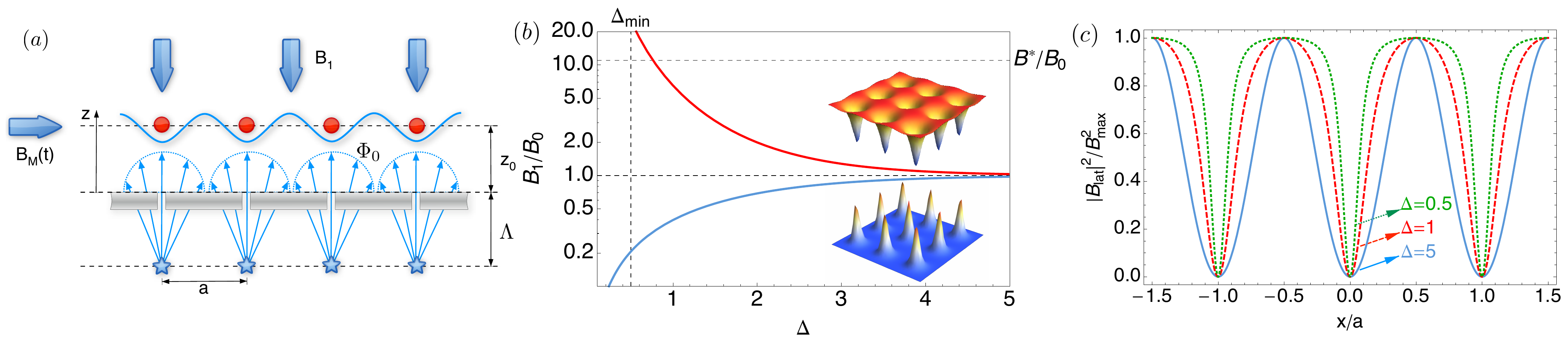}
\caption{ (Color online) (a) Schematic illustration of the superconducting vortex lattice for ultracold atoms. (b) Applied field  $B_{1}$ as a function of $\Delta$. Inset: $|\BB_\text{lat}(x,y,z_{0})|^2$ for a square lattice, $\Delta=2$, and $B_{1}>B_{0}$ (deep wells) and $B_{1}< B_{0}$ (sharp peaks). (c) $ |\BB_\text{lat}(x,0,z_{0}|^{2}$ in units of $B^{2}_\text{max}\equiv |\BB_\text{lat}(x=a/2,0,z_{0})|^{2}$ as a function of $x/a$ for a square lattice for different $\Delta$.   }
\label{Fig1}
\end{center}
\end{figure*}

Our proposal consists in using the magnetic field created by the vortex lattice above the film $\BB_\text{V}(\rr,z>0)$ to trap neutral atoms in a two-dimensional magnetic nanolattice with a geometry dictated by the nanoengineered antidot lattice.  One can use well-known results in the field of superconductivity to approximate the field above the thin film as the one generated by an array of magnetic charges of strength $2 \Phi_{0}$ situated at a distance $z=-\Lambda$ below the film, see Fig.~\ref{Fig1}a and SM.  This leads to
\be \label{eq:Banalytic}
\begin{split}
B^{x(y)}_\text{V}(\rr,z) &\approx   B_{0} e^{-\Delta_{z}} g_{x(y)}(\rr,z),\\
B^{z}_\text{V}(\rr,z) &\approx   B_{0} \left[1+ e^{-\Delta_{z}}  g_{z}(\rr,z)  \right].
 \end{split}
\ee
We have defined $B_{0} \equiv  \Phi_{0}/a^{2}$, $\Delta_{z} \equiv 2 k (z+\Lambda)> 2 k \Lambda  \equiv \Delta_\text{min}$, $k \equiv  \pi /a$, and
\be
\begin{split}
g_{x(y)}(\rr,z>0) &\equiv \sum_{\KK \neq \bold{0}} \frac{K_{x(y)}}{|\KK|}  \sin ( \KK \cdot \rr ) e^{  \Delta_{z}(1-|\KK|/k) }, \\
g_{z}(\rr,z>0) & \equiv \sum_{\KK \neq \bold{0}}   \cos ( \KK \cdot \rr ) e^{  \Delta_{z}(1-|\KK|/k) }, 
\end{split}
\ee
where the sum is over all reciprocal lattice vectors $\KK$. 
The $x$ and $y$ components of $\BB_\text{V}$ are equal to zero on top of the vortices, namely at $\rr=\RR$. The $z$ component is always positive and tends to an homogeneous field of strength $B_{0}$ at long distance from the surface due to the infinite extension of the plane. General nonperiodic structures obtained by nanoengineering can also be considered.

Alkali metal atoms in low fields of strength $ \lesssim 30$~mT, where the Zeeman shift of hyperfine levels is linear, experience a potential of the form $V_\text{lat}(\rr,z) = \mu_{m_{F}} |\BB(\rr,z)|$~\cite{Folman02, Fortagh07}. The local field interacting with the atoms  is denoted by $\BB$ and will be composed of the one generated by the vortices $\BB_\text{V}$ plus additional bias fields, see below. The magnetic dipole moment is given by $\mu_{m_{F}} \equiv m_{F} g_{F} \mu_\text{B}$, where $m_{F}$ is the magnetic quantum number, $g_{F}$ is the Land\'e $g$-factor, and the positive number $\mu_\text{B}$ is the Bohr magneton. Thus, low-field-seeking states $g_{F} m_{F} > 0$ can be trapped at the local minima of $ |\BB(\rr,z)|$~\cite{Folman02, Fortagh07}. 

\begin{figure}
\begin{center}
\includegraphics[width=  0.8  \linewidth]{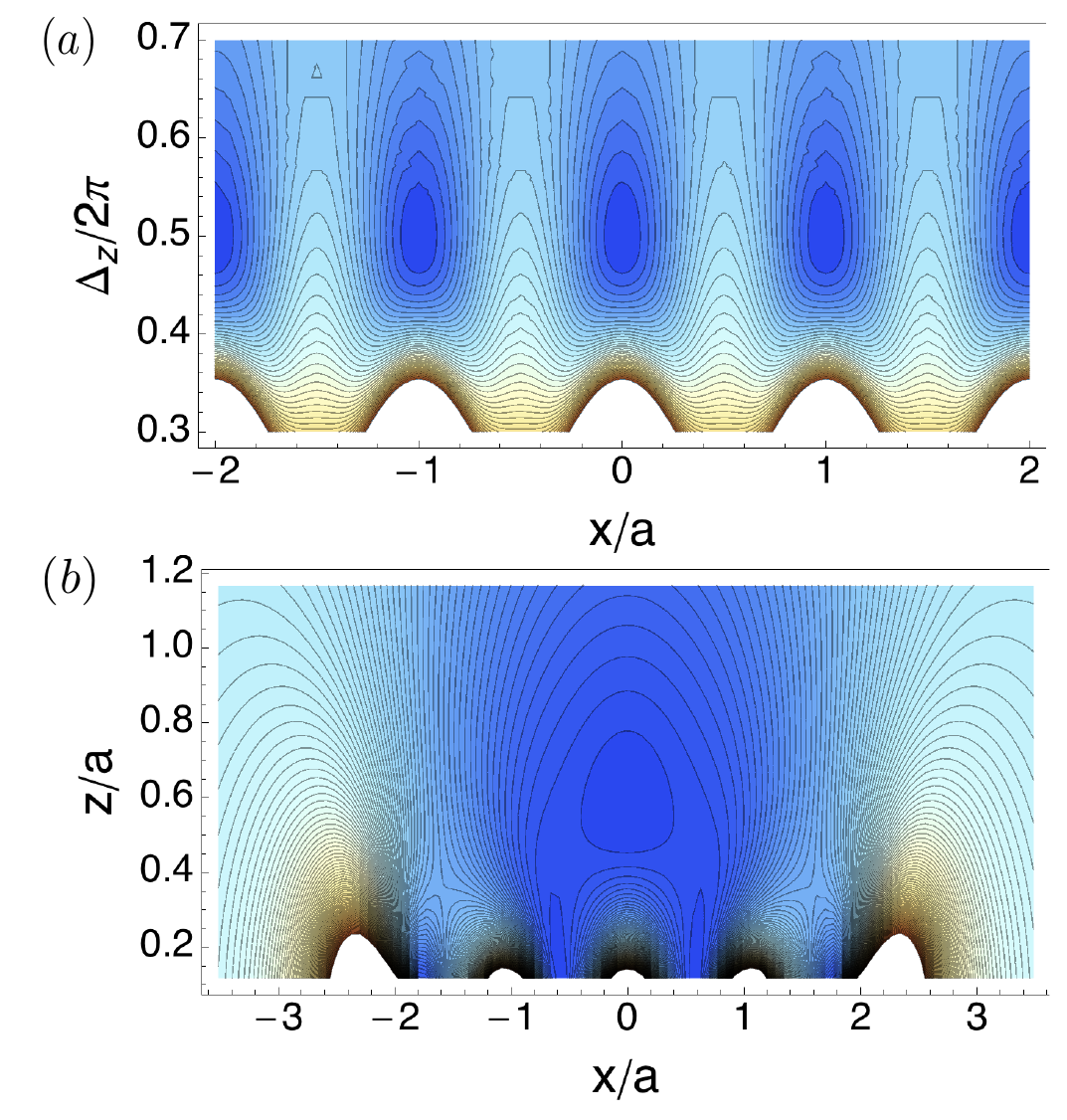}
\caption{ (Color online)  (a) A contour plot of $|\BB_\text{lat}|^{2}$ at $y=0$ is shown for an infinite square lattice as a function of $x/a$ and $\Delta_{z}$ for $\Delta= \pi$  using \eqcite{eq:Banalytic}. (b) The same plot as (a) is shown with a set of $3 \times 3$ localized vortices located on a squared lattice in a thin superconductor (the borders of the plane are at $x/a=\pm 2.45$) by numerically solving the London equation\cite{Brandt05b} and using $\Lambda \approx a$. }
\label{Fig2}
\end{center}
\end{figure}

The field generated by the vortices does not have local minima of $|\BB_\text{V}|$ since the $z$-component is always positive. For this reason,  we propose to add   a perpendicular bias field of the form $\BB_{1}=B_{1}(0,0,-1)$ and define $\BB_\text{lat}(\rr,z)=\BB_\text{V}(\rr,z)+\BB_{1}$. The field $\BB_{1}$ can be considered homogeneous even close to the surface provided that a thin film of $d \lesssim \lambda$ is used. We have validated this assumption by numerically calculating the field distribution and induced currents in a superconducting disk of finite radius and thickness using an energy minimization procedure \cite{Sanchez01,Chen08}, see SM.  The strength of $B_{1}$ is limited by the fact that it should not induce extra vortices. This leads to the condition $B_{1} < B^\star + \min_{\rr} B^{z}_\text{V}(\rr,0)$, where $B^{\star} \approx \Phi_{0}/(4\pi \Lambda^{2} )$. The local minima of $|\BB_\text{lat}(\rr,z)|$ are obtained on top of the vortices, $\rr = \RR$, when $B_{1}>B_{0}$, at a position $z_{0}$ given by the solution of the equation  $B^{z}_\text{V}(\RR,z_{0})=B_{1} $, see Fig.~\ref{Fig1}b, where we have defined $\Delta \equiv \Delta_{z_{0}}$. In this case, the conditions $B^{\star}> B_{1} > B_{0}$ can be fulfilled provided $\Lambda \gtrsim a$. In Fig.~\ref{Fig2}a we plot $|\BB_\text{lat}|^{2}$ (as the final form the potential depends on the modulus square, see \eqcite{eq:Vlatt1}) in the $x-z$ plane at $y=0$. As a validation of the model, we also plot the same quantity in Fig.~\ref{Fig2}b for an array of $3\times 3$ vortices in a finite plane numerically solving the London equation using the method presented in~\cite{Brandt05b}. One can observe that there is also a minimum of $|\BB_\text{lat}(\rr,z)|^{2}$ on top of the central vortex.
In the case $B_{1}<B_{0}$, one confines atoms between vortices with a much shallower trap, see Fig.~\ref{Fig1}b.

Majorana losses~\cite{Brink1997}, namely spontaneous spin flips rendering the state of the atom into a high-field seeker, occur when $|\BB| \sim 0$. Since the minima of $|\BB_\text{lat}|$ correspond to zero field, we suggest to use an effective time-averaged, orbiting potential~\cite{Petrich95} generated by adding to $\BB_\text{lat}$  an homogeneous time-dependent field  parallel to the thin film 
$\BB_\text{M}(\text{t})=B_\text{M} (\sin \omega_\text{M} \text{t}, \cos \omega_\text{M} \text{t}, 0)$.
By time averaging we have that $\avg{\BB_\text{M}(\text{t})}= {\bf 0}$, but $\avg{|\BB_\text{M}(\text{t})|} = B_\text{M}$. Assuming $B_\text{M} \gg \max |\BB_\text{lat}(\rr,z_{0})| $ and considering that the total field experienced by the atoms is given by $\BB=\BB_\text{lat} + \BB_\text{M}$, we have 
$\avg{|\BB|} \approx B_{M} + |\BB_\text{lat}|^{2}/(2 B_\text{M})$, which does not contain zero-field local minima. Using 
$\omega_\text{t} \ll \omega_\text{M} \ll \omega_\text{L} \equiv \mu_{m_{F}} B_\text{M} /\hbar$, the effective time-averaged magnetic potential for the atoms is given by~\cite{Petrich95}
\be \label{eq:Vlatt1}
V_\LAT(\rr,z)  \approx \hbar \omega_\text{L} + \frac{\mu_{m_{F}} }{2 B_\text{M}} |\BB_\text{lat}(\rr,z)|^{2}.
\ee
This potential depends on $|\BB_\text{lat}|^{2}$,  has nonzero field minima reducing Majorana losses to a rate given by $\Gamma_\text{ML}/2 \pi \approx \omega_\text{t} \exp[- 4 \omega_\text{L}/\omega_\text{t}]$~\cite{Brink1997}, and confines atoms on a magnetic lattice whose geometry is dictated by the vortex lattice. Note that the bias field $B_{1}$ can be used to control the trapping height, see Fig.~\ref{Fig1}b. This might be used to adiabatically load the ultracold atoms into the magnetic lattice from an external dimple trap~\cite{StamperKurn98}.

\begin{figure}
\begin{center}
\includegraphics[width= .8 \linewidth]{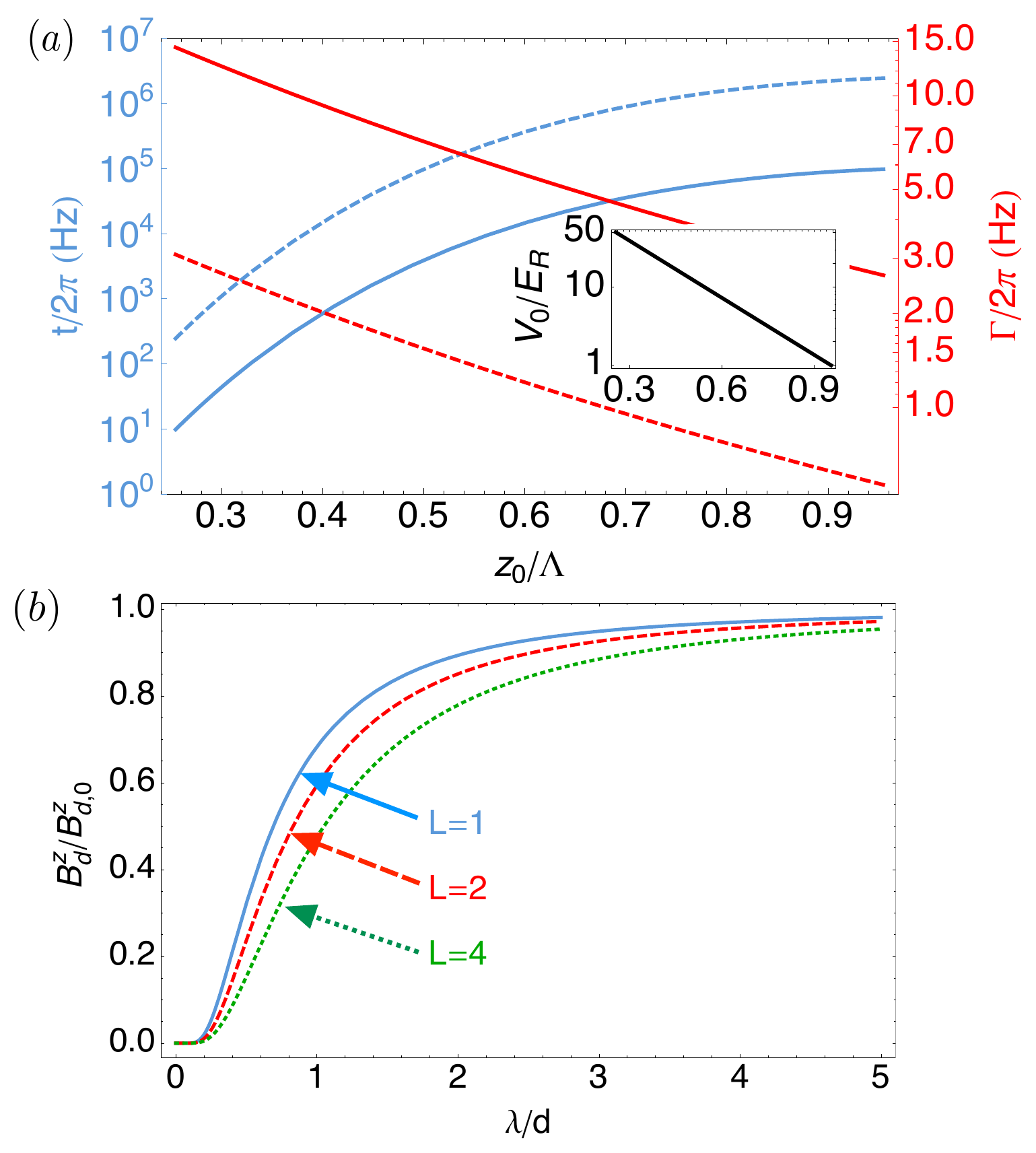}
\caption{(Color online)  (a) Tunneling rate $t$ (left axis), decoherence rate (right axis) $\Gamma=\Gamma_\text{ML}+\Gamma_\text{sf} + \Gamma_{0 \rightarrow 1}$, and $V_{0}/E_\text{R}$ (inset) as a function of the trapping position $z_{0}$ in units of $\Lambda$ for a square lattice at sufficiently large $\Delta$. The range of the plot is limited to the values of $z_{0}/\Lambda$ for which $50 > V_{0}/E_\text{R}>1$. We used the atomic parameters of lithium,   $a=3 \Lambda/2$, $\Lambda=100$ nm (solid lines) and $\Lambda=20$ nm (dahsed lines), $B_\text{M}$ such that $\omega_\text{ML}=5 \omega_\text{T}$, and $\eta/\Lambda=10^{-7} $~$\text{Kg}/(\text{s m})$~\cite{Pompeo08}. (b) The ratio $B^{z}_\text{d}/B^{z}_\text{d,0}$ as a function of $\lambda/d$ for $L=1,2,4$.   
}
\label{Fig3}
\end{center}
\end{figure}

The strength of the Hubbard parameters and the dependence with the physical parameters of the superconducting vortex lattice (SVL) proposed here can be obtained in analogy to OLs~\cite{Bloch08a}. In particular, let us consider a square lattice with $B_{1}>B_{0}$ and $\Delta$ sufficiently large, such that the potential \eqcite{eq:Vlatt1} can be approximated to $V_\LAT(\rr,z)\approx V_{0} [ \sin^{2}( k x) + \sin^2 (k y)] $, where $V_{0} \equiv  8 B^{2}_{0} \exp[  -2 \Delta]  \mu_{m_{F}}/B_\text{M}$. This potential has the same form as the typical one obtained in OLs. The distance between the trapped atoms in the SVL is given by $a$, therefore  $2 a$ plays the role of the optical wavelength in OLs. The role of the laser intensity in OLs, that can be used to modulate the trap depth, is taken in the SVL by the strength of the bias field $B_{1}$. Recall that $\Delta$ depends on $B_{1}$, see Fig.~\ref{Fig1}b. 
In Fig.~\ref{Fig3}a, we plot the tunneling rate $t \approx 4 E_\text{R} \left(V_0/E_\text{R} \right)^{3/4}  \exp \left[-2 (V_0/E_\text{R} )^{1/2} \right]/\sqrt{\pi}$~\cite{Bloch08a} as a function of  $z_{0}/\Lambda$ for Li  (using $a=3 \Lambda/2$ for $\Lambda=100$ nm and $\Lambda=20$ nm) and compare it with the decoherence rates that we discuss below. This is plotted in the range $50>V_{0}/E_\text{R}>1$ (see inset of Fig.~\ref{Fig3}a), which includes the tight-binding regime where $U \gg t$~\cite{Bloch08a}. The on-site repulsion $U < \hbar \omega_\text{t} \sim 2 E_\text{R} (V_{0}/E_\text{R})^{1/2}$ can be calculated solving the ultracold two-body collisional problem in a tight harmonic trap~\cite{Bolda02, Gullans12}. Comparing to OLs, the case $\Lambda=20$ nm leads to more than two orders of magnitude larger values of $E_\text{R}$, and therefore, significantly less stringent low-entropy requirements for the simulation of quantum magnetism in Hubbard Hamiltonians~\cite{Greif12}, as discussed in the introduction.
In contrast to OLs, the potential \eqcite{eq:Vlatt1} also permits to design dense lattices with higher Fourier components. Here, by reducing $\Delta$, more reciprocal vectors of different frequencies enter into play, see Fig.~\ref{Fig1}c. For $B_{1}<B_{0}$ this can lead to interesting potentials with sharp repulsive structures, see the inset of Fig.~\ref{Fig1}b.

Atoms in magnetic traps are subjected to decoherence (spin flips and motional heating) due to magnetic field fluctuations at the Larmor frequency $\omega_\text{L}$~\cite{Henkel99}. In metal surfaces, these fluctuations are generated by thermally excited motion of electrons (Johnson noise)~\cite{Henkel03}. Superconducting vortex-free surfaces have been predicted to dramatically reduce Johnson noise by 6-12 orders of magnitude~\cite{Skagerstam06,Hohenester07,Skagerstam07}.  However, experiments in superconducting atom chips have only shown  a moderate improvement since they operate in a regime where uncontrolled superconducting vortices are present~\cite{Hufnagel09, Kasch10, Nogues09, Fruchtman12}. In the SVL proposed here, it is clear that a source of magnetic field fluctuations will be given by the thermal jiggling of the pinned vortices. In the SM we use a standard phenomenological model of vortex dynamics~(see \cite{Pompeo08} and references therein) in order to estimate the spin-flip rate $\Gamma_\text{sf}$ and the motional heating rate $\Gamma_{0 \rightarrow 1}$~\cite{Henkel99} induced by the thermal motion of vortices. They are given   (up to some constant factors) by
\be
\begin{split}
\Gamma_\text{sf} &\sim 3 \pi^{3}  \frac{\mu_{m_{F}}^{2}}{\hbar^{2}}\frac{ k_\text{B}T}{k_\text{p}  } \frac{\omega_\text{d}}{\omega_\text{L}^{2}+\omega_\text{d}^{2}}\frac{B^{2}_{0}}{a^{2} \Delta^{4}},  \\
\Gamma_{0 \rightarrow 1} & \sim (2 \pi)^{5} \frac{ \mu_{m_{F}}^{2} }{\hbar^{2}} \frac{  k_\text{B}T}{ k_\text{p}   } \frac{\omega_\text{d}}{\omega_\text{t}^{2}+\omega_\text{d}^{2}}     \frac{x_{0}^{2} B^{2}_{0} }{a^{4} \Delta^{6}}.
\end{split}
\ee
Here $x_{0}=\sqrt{\hbar/(2 m_\text{a}\omega_\text{t})}$, $k_\text{p}=  \Phi_{0}^{2}/(2 \mu_{0} \Lambda a^{2})$ (provided $2 \pi \Lambda \gg a$) is the spring constant given by the repulsive force with the lattice, see~\cite{Brandt09}, and  $\omega_\text{d}=k_\text{p}/\eta$ with $\eta$ being the vortex viscosity coefficient is the so-called depinning frequency, which marks the crossover between elastic motion, dominant at lower frequencies, and purely dissipative motion, arising at higher frequencies~\cite{Pompeo08}. Using typical numbers for the vortex viscosity, these rates are remarkably small compared to the tunneling rate in Hubbard Hamiltonians, see Fig.~\ref{Fig3}a. 

Other sources of decoherence and practical considerations might be relevant in the eventual experiment. For instance, as analyzed in the SM,  the position of the vortices has to be very accurate, with an error less than $1- 2$ $\%$ in the distance between them. Otherwise, the trap depths and thereby the tunneling rates will fluctuate throughout the lattice. This can constitute a serious challenge in the nanofabrication of regular antidot lattices. In this respect, using triangular lattices spontaneously formed in a film without artificial pinning might be advantageous. The randomness in size and shapes of the antidots might also lead to imperfections, nevertheless, the flux in each vortex is given by the constant of nature $\Phi_{0}$. 
Time=dependent fields might induce dissipation in the vortices. As discussed before, we apply a time-dependent field $\BB_\text{M}(t)$ to avoid Majorana losses. This field is parallel to the film, and in the ideal case, would not interact with the vortices. In any case, to reduce dissipation it will be convenient to use $\omega_\text{M} \ll \omega_\text{d}$.

Let us discuss the possibility to perform magnetic local addressing of the atoms in the lattice.
 We propose to place a magnetic tip close to the bottom side of the film, at position $z=-d-a_\text{d}$, to locally interact with the atoms trapped above. In the SM, we obtain the analytical expression of the magnetic field above the film for a given thickness $d$ and London penetration depth $\lambda$. The ratio between the magnetic field $B^{z}_\text{d}$ at $z=z_{0}+d$  in the presence of a superconducting film with London penetration depth $\lambda$, with the corresponding one in case of not having the film $B^{z}_\text{d,0}$, depends on the dimensionless parameters $\lambda/d$ and $L\equiv (z_{0}+a_\text{d})/d$. In Fig.~\ref{Fig3}b, this ratio is plotted as a function of $\lambda/d$ for different $L$. Note that even for $\lambda=d$, $B^{z}_\text{d} \sim B^{z}_\text{d,0}/2$ for $L=2$. As discussed in~\cite{Wei96}, the minimum distance for which the dipole will create a vortex is given by 
$a_{1}= \Lambda \sqrt{\mu_\text{0} m_\text{d} /[ \Phi_{0} \Lambda \ln (\Lambda/\xi)]}$,
this equation is valid for $a_{1} > \Lambda$. With typical numbers ($\Lambda=100$ nm and $\xi=10$ nm), $a_{1} \sim 2 \Lambda$ for a dipole of magnetic moment $m_d \sim 10^{8} \mu_\text{B} $.  Using the maximum magnetic moment and  $\lambda=d=\Lambda=z_{0}=a_\text{d}/2=100 $ nm (note that $L=3$) this leads to a coupling to the atom of $g_\text{d} \sim \mu_\text{B} B^{z}_\text{d}/\hbarÊ\sim 0.4  \mu_\text{B} B^{z}_\text{d,0}/\hbar  \sim 2\pi  \times 10^{8}$~Hz. Since the neighbor atoms are farther away from the dipole, the coupling is reduced at least by a factor of $2$, which permits to perform local addressing by adding different phases into the internal state of the atoms.  In order to measure the collective state of the atoms one can release them from the trap and perform time-of-flight measurements~\cite{Bloch08a}.

To conclude, we have shown that the control and manipulation of superconducting vortices in thin films can be used to trap neutral atoms in a dense lattice near a surface in a superconducting state, whose macroscopic coherence leads to promising scalings regarding surface decoherence. 
 The interplay between superconductivity and atomic physics, as proposed here, might pave the way towards all-magnetic schemes for quantum computation and simulation with neutral atoms. Moreover, this hints at the possibility of using ultracold atoms to probe important properties of high-$T_\text{c}$ superconductivity.

 We acknowledge funding from  the EU projects AQUTE and ITN Coherence, Spanish Consolider NANOSELECT (CSD2007-00041) and MAT2012-35370 projects, and the Austrian Science Fund (FWF) through SFB FOQUS. P.Z. thanks  the hospitality at MPQ as a MPQ fellow.

\newpage

\newpage

\section*{SUPPLEMENTARY INFORMATION}

\section{Superconducting Vortex Lattice}

In this section we provide more information about the superconducting vortex lattice. In particular, we review some known results in the field of superconductivity for the reader more specialized in atomic physics and quantum simulation.  In particular, in Sec.~\ref{SS:PV} we derive the magnetic field generated by a single vortex in a thin film of $d \ll \lambda$ using the London approximation $\lambda \gg \xi$. We also discuss the approximation used in the main text consisting of replacing the magnetic field generated by a vortex with the one generated by a magnetic monopole. In Sec.~\ref{SS:VV} we discuss the interaction of a vortex with an antidot as well as the vortex-vortex interaction also within the London approximation.

\subsection{Pearl vortex and monopole approximation} \label{SS:PV}

We consider a thin film of thickness $d$ whose upper side is at the plane $x-y$ with $z=0$. In the following, we calculate the magnetic field above the film created by a single vortex situated at $\RR$, namely $\BB_\RR (\rr,z>0)$. In the London limit $\lambda \gg \xi$, the analytical solution for any thickness $d$ can be found in~\cite{Carneiro00}. Here, we will recall the particular case of $d\gg \lambda$, originally derived by Pearl~\cite{Pearl64}. 

We define the effective London penetration $\Lambda\equiv \lambda^{2}/d$ and use standard cylindrical coordinates
$(\rho,\phi,z)$. For small $d/\lambda$, the current density ${\bf J}$ inside the
superconducting film is basically uniform across the thickness and can be
regarded as a current sheet ${\bf K}(x,y)\simeq \int_{-d}^0 {\bf J}(x,y,z) {\rm
d}z$. We consider the
vortex placed in the origin of coordinates and we will drop the subindex $\RR$ to ease the notation. In the thin film
limit case, currents can be obtained from the discontinuity of the radial
component of the magnetic induction field ${\bf B}$. Because of the
symmetry, ${\bf K}$ has only angular
component given by ($\mu_0$ is the vacuum permeability)
\be
\label{eq.kt}
\mu_0 K_\phi(\rho) =  B_\rho(\rho,z\rightarrow 0^+)- B_\rho(\rho,z\rightarrow
0^-) 
\ee
Outside the superconducting region, the vector potential ${\bf A}$  must satisfy
$\nabla \times
(\nabla \times {\bf A}) = 0$, since $\nabla \times {\bf B}=0$. Together with the
choice of Gauge $\nabla \cdot {\bf A} =0$, and the consideration of cylindrical symmetry and mirror
symmetry with respect to the $z=0$ plane, this yields the known general
solution
\be
\label{eq.at}
A_\phi (\rho, z) = \int_0^\infty C(k) J_1(k\rho) e^{-k|z|} {\rm d}k,
\ee
where $J_i()$ is the Bessel function of the first kind and $i$th order. From
Eqs. (\ref{eq.kt}) and (\ref{eq.at}), we have
\be
\label{eq.muk}
\mu_0 K_\phi(\rho)= \int_0^\infty 2 \, C(k) \, k \, J_1(k\rho) {\rm d}k.
\ee
To evaluate $C(k)$ we consider the superconducting film as described with the
London equation in which a vortex with coherence length $\xi$, much smaller than
any other dimension of the problem, is placed at the origin. London equation
reads~\cite{DegennesBook,Pearl64}
\be
\nabla \times {\bf K} = - \frac{1}{\mu_0 \Lambda} {\bf B} +
\frac{\Phi_0}{\Lambda \mu_0} \frac{\delta(\rho)}{2\pi\rho} \hat{{\bf z}},
\ee
where $\delta()$ is the Dirac's delta function. This equation can also be
obtained from the fluxoid quantization condition~\cite{Clem91}. Using \eqcite{eq.muk}, considering the continuity of the $z$ component of ${\bf B}$, and
making use of Hankel transformations we obtain
\be
C(k) = \frac{\Phi_0}{2\pi (2 k \Lambda +1)}.
\ee
Thus, the sheet current density, and the magnetic induction can be
obtained in all space. In particular, for the thin film case and for $z>0$, we
have
\be
\begin{split}
 \label{B:PV}
 B_\rho(\rho,z>0)&=\int_0^\infty \frac{\Phi_0}{2\pi} \frac{k}{2
k \Lambda +1} J_1(k\rho) e^{-k z}, \\
 B_z(\rho,z>0)&=\int_0^\infty \frac{\Phi_0}{2\pi} \frac{k}{2 k
\Lambda +1}
J_0(k\rho) e^{-k z}.
\end{split}
\ee
For $z<0$ we could use the mirror symmetry with respect the $z=0$ plane.

\begin{figure}[tb]
\begin{center}
\includegraphics[width= 0.8 \linewidth]{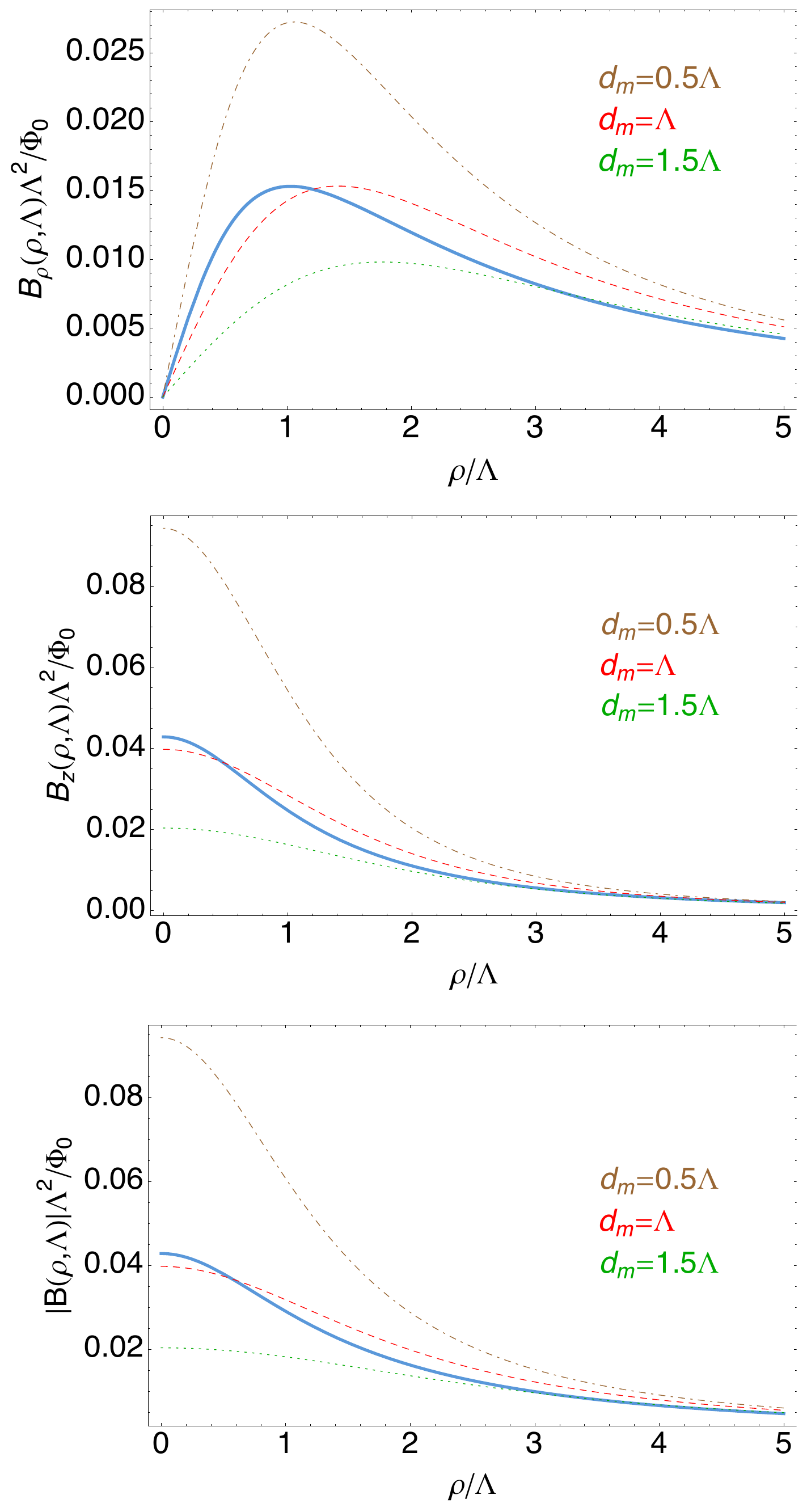}
\caption{Comparison of the field components and the modulus using the exact formula of the Pearl vortex (solid blue line), see \eqcite{B:PV}, and the approximation of the monopole,  see \eqcite{B:M}, for $d_\text{m}=0.5\Lambda,\Lambda,1.5 \Lambda$ (brown dot-dashed, red dashed, green dotted). }
\label{f.pv}
\end{center}
\end{figure}

As discussed in~\cite{Clem91} and specially in~\cite{Carneiro00}, the field generated by a vortex can be approximated by the one created by a monopole of charge $2\Phi_{0}$ placed at $z=-d_\text{m}$, that is,
\be
\begin{split}
 \label{B:M}
 B^{0}_\rho(\rho,z>0)&=\int_0^\infty \frac{\Phi_0}{2\pi} k J_1(k\rho) e^{-k (z+d_\text{m})}\\
 &= \frac{\Phi_0}{2\pi}  \frac{\rho}{\left[ \rho^{2} + (z+d_\text{m})^{2} \right]^{3/2}}
, \\
 B^{0}_z(\rho,z>0)&=\int_0^\infty \frac{\Phi_0}{2\pi} k  J_0(k\rho) e^{-k (z+d_\text{m})} \\
&= \frac{\Phi_0}{2\pi}  \frac{z+d_\text{m}}{\left[ \rho^{2} + (z+d_\text{m})^{2} \right]^{3/2}}.
\end{split}
\ee
In Fig.~\ref{f.pv}, we compare the components and the modulus of the magnetic field along $\rho$ at $z=\Lambda$ using \eqcite{B:PV} and \eqcite{B:M} for different $d_\text{m}$. A good approximation for the near field is given by  $d_\text{m}=\Lambda$. See also~\cite{Carneiro00}, where the field lines for both cases are plotted showing very clearly that this simple approximation is indeed very good.

\subsection{Vortex-vortex and vortex-antidot interaction} \label{SS:VV}

\begin{figure}
\begin{center}
\includegraphics[width=\linewidth]{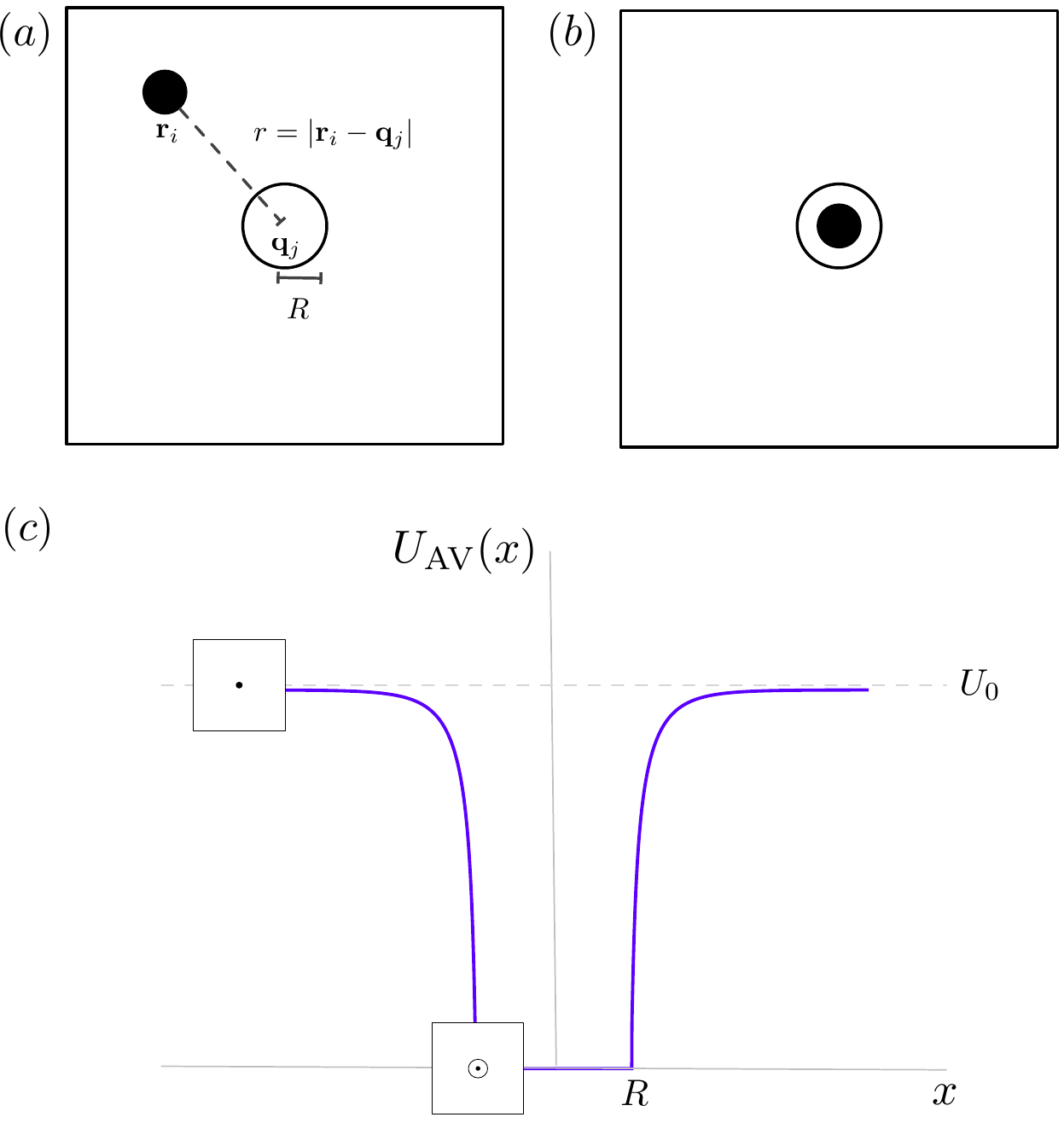}
\caption{(a) A vortex is attracted to an antidot of radius $R$. The center of the vortex is at a distance $r$ to the center of the antidot. (b) The vortex is pinned by the antidot. (c) Antidot pinning potential.}
\label{Fig1}
\end{center}
\end{figure}

The interaction of a vortex (V) at position $\rr_{i}$  with an antidot (AD) at position $\qq_{j}$ of radius $R$ has been analytically studied in the London approximation ($\lambda \gg \xi$)~\cite{Mkrtchyan72,Buzdin96,Nordborg00}. This allows to compute the interaction energy of a vortex separated a distance $r\equiv |\rr_{i}-\qq_{j}| $ to the center of the antidot, see Fig.~\ref{Fig1}. The interaction potential is found to have the form of Fig.~\ref{Fig1}c, which resembles a finite square well. The depth of the well $U_{0}$ is the difference between the energy of an isolated vortex $E_\text{V}$ minus the energy of a microhole of radius $R$ with one quantum of magnetic flux, $E_\text{AD}$. That is,
\be
U_{0}=E_\text{V} - E_\text{AD}
\ee
The length of the region where the potential is zero is $2R$.

References \cite{Mkrtchyan72,Buzdin96} assume $R,r \ll \lambda$ , which permits a useful analogy to electrostatics in order to use the method of images. Reference~\cite{Nordborg00} solves the problem for any value of $R$, such that in the limit $R\ll \lambda$ they recover the results of  \cite{Mkrtchyan72,Buzdin96}, and, in the limit of very large $R$, they recover the well-known Bean-Livingston formula for the interaction between a vortex and an insulating barrier. The results of these references are valid for Abrikosov vortices. For Pearl vortices, (see Eqs.~(3.26) and (3.27) in \cite{Wei96}), one obtains an expression for the self-energy of a Pearl vortex (PV) as well as the self-energy of an AD with one quanta of the magnetic flux,
\be
\begin{split}
E_\text{PV}&=\frac{\Phi_{0}^{2}}{4 \pi \mu_{0} \Lambda}  \ln \frac{2 \Lambda}{\xi}\\
 E_\text{AD}&=\frac{\Phi_{0}^{2}}{4 \pi \mu_{0} \Lambda} \ln \frac{2 \Lambda}{R}.
\end{split}
\ee
Thus, 
\be \label{eq:V0}
U_{0}=E_\text{PV}- E_\text{AD} =\frac{\Phi_{0}^{2}}{4 \pi \mu_{0} \Lambda} \ln \frac{R}{\xi}.
\ee
This energy will be positive provided $R>\xi$. Using $R=5 \xi$ and $\Lambda=100$~nm, one obtains that $U_{0}/k_\text{B} \approx 3 \times 10^{5}$~K.

In reference \cite{Nordborg00} (see Eq.~(30) and (31) in this reference) it is also provided  the analytical expression of the total energy, including the interaction.  Indeed, we used the exact formula to plot Fig.~\ref{Fig1}. In the limit $R < |\rr_{i}-\qq_{j}| \ll \Lambda$, the interaction energy reads
\be
U_\text{AV}(|\rr_{i}-\qq_{j}|)=U_{0} \ln \left[ 1- \frac{R^{2}}{|\rr_{i}-\qq_{j}|^{2}}\right].  
\ee
In general, one can approximate the potential as a finite square well, namely
\be \label{eq:AV}
U_\text{AV}(|\rr_{i} - \qq_{j}|) \approx U_{0} \Theta \left( |\rr_{i} - \qq_{j}| >R  \right),
\ee
where $U_{0}=\Phi_{0}^{2}/(4 \pi \mu_{0} \Lambda) \ln ( R/\xi)$ and $\Theta (x)$ is the Heaviside function. In the following, we show that the interaction between vortices in a pinned vortex lattice yields an additional pinning force. 

The interaction between Pearl vortices is nicely studied by Brandt in~\cite{Brandt09}. The thin-film problem differs from the behaviour of currents and vortices in bulk superconductors by the dominating role of the magnetic stray field outside the film. The interaction between the vortices occurs mainly by this stray field and thus has very long range (interacting over the entire film width). In bulk superconductors the vortex currents and the vortex interaction are screened, and thus decrease exponentially over the length $\lambda$.

The interaction potential between two PV (separated by a distance $r_{ij}\equiv |\rr_{i}-\rr_{j}|$) is given by~\cite{Brandt09}:
\be \label{eq:VVpotential}
\begin{split}
U_\text{VV}(r_{ij})&= \frac{\Phi^{2}_{0}}{\mu_{0}} \int \frac{d^{2} k}{ 4 \pi^{2}} \frac{2 \cos (\kk \cdot (\rr_{i}-\rr_{j}))}{k+2 \Lambda k^{2}} \\
&= \frac{\Phi^{2}_{0}}{\mu_{0}} \int_{0}^{\infty} \frac{d k}{ 2 \pi} \frac{2 J_{0} (k r_{ij})}{1+2 \Lambda k}.
\end{split}
\ee
This expression can be written in terms of Struve and Weber functions,
\be
U_\text{VV}(r_{ij}) =  \frac{\Phi^{2}_{0}}{4 \Lambda \mu_{0}} \left[H_{0} \left(\frac{r_{ij}}{2 \Lambda} \right) +Y_{0} \left(\frac{r_{ij}}{2 \Lambda} \right) \right],
\ee
and can be simplified in the following limits
\be
U_\text{VV}(r_{ij}) \approx   \frac{\Phi_{0}^{2}}{\mu_{0} \pi} \frac{ \ln ( 2.27 \Lambda/r_{ij} )}{2 \Lambda} \hspace{1em}  \text{for} \hspace{1em} r_{ij}\ll \Lambda,
\ee
and
\be
U_\text{VV}(r_{ij}) \approx   \frac{\Phi_{0}^{2}}{\mu_{0} \pi r_{ij}} \hspace{1em}  \text{for} \hspace{1em} r_{ij} \gg \Lambda.
\ee
In the long-distance limit the PVs interact as point-like charges.


We consider the interplay between the vortex-vortex interaction~\eqcite{eq:VVpotential} and the interaction between a vortex and an antidot~\eqcite{eq:AV}. For instance, how near can  two antidots of radius $R$ be such that 2 vortices can still be pinned? This can be easily answered by solving the equation $U_\text{VV}(r_{ij})Ê\approx U_{0}$, which for $R=10 \xi$ leads to $r_{ij} \approx \Lambda$. This indicates that the antidots should be separated a distance $\gtrsim \Lambda$ so that the repulsion between vortices does not prevent them from sitting at the pinning sites.

Let us now calculate the energy due to the interaction between vortices 
assuming that the the vortices are sitting at ideal-lattice points $\RR$ with reciprocal lattice vectors $\KK$. This is given by~\cite{Brandt09}
\be
\begin{split}
U_\text{int}(\{ \RR \})&=\frac{1}{2} \sum_{\RR\neq \RR'} U_\text{VV}(|\RR-\RR'|)\\
&= \frac{1}{2} \frac{\Phi^{2}_{0}}{\mu_{0}} \sum_{\RR\neq \RR'} \int  \frac{d^{2} k}{ 4 \pi^{2}}  \frac{2 \cos ( \kk \cdot (\RR-\RR') )}{k+2 \Lambda k^{2}} 
\end{split}
\ee
The infinite lattice sum (over all lattice sites) can be evaluated using
\be
\sum_{\RR} \exp[\im \kk \cdot \RR] = \frac{4 \pi^{2}}{a^{2}}  \sum_{\KK} \delta_{2}(\kk-\KK),
\ee
where $1/a^{2}$ is the density of lattice points $\RR$; the density of reciprocal-lattice points is $a^{2}/(4 \pi^{2} )$. One obtains
\be
\begin{split}
&\frac{U_\text{int}(\{ \RR \})}{N}= \frac{\Phi^{2}_{0}  }{\mu_{0} a^{2}} \times \\
& \left[ \sum_{\KK \neq 0} \frac{1 }{K+2 \Lambda K^{2}} - \int_{0}^{\infty}  \frac{d k}{ 2 \pi n}  \frac{k}{k+2 \Lambda k^{2}} \right]
\end{split}
\ee
With the same formalism, we can readily obtain the interaction energy between a vortex that has been slightly displaced by the distance $\rr$ of its lattice site $\RR=0$ with the rest of the lattice. This interaction energy $U_\text{pV}(\rr)$ is given by~\cite{Brandt09}
\be
\begin{split}
&U_\text{pV}(\rr)= \frac{\Phi^{2}_{0}  }{\mu_{0} a^{2}} \times \\
& \left[ \sum_{\KK \neq 0} \frac{2 \cos (\KK \cdot \rr) }{K+2 \Lambda K^{2}} - \int \frac{d^{2} k}{ 4 \pi^{2} n}  \frac{2 \cos (\KK \cdot \rr)}{k+2 \Lambda k^{2}} \right].
\end{split}
\ee
This can be used to obtain the pinning force of a vortex in the lattice due to the interaction with the rest. That is, one can obtain $U_\text{pV}(\rr) \approx U_\text{pV}(0) + U_\text{pV}''(0) r^{2}/2$. Assuming $2 \pi \Lambda \gg a$, one gets that the pinning force has a restoring spring constant given by~\cite{Brandt09}
\be \label{eq:kp}
k_\text{p}= U_\text{pV}''(0)= \frac{\Phi_{0}^{2}}{2 \mu_{0} \Lambda a^{2}}.
\ee
(see Eq.~(14) in \cite{Brandt09}).

Given a pinning lattice with some separation distance between ADs, reference~\cite{Pogosov03} studies the transition between vortices sitting into the pinning lattice to the triangular lattice obtained when interaction between PVs dominates. Given a certain configuration of ADs and a certain number of PVs we could also numerically minimize the total energy in order to obtain the final location of PVs. These problems have been studied in~\cite{Pogosov03} and references therein. 
 The vortex lattice can be prepared by  cooling the film in the presence of an homogeneous perpendicular matching field of strength $B_\text{a}=\Phi_{0}/a^{2}$ such that a commensurate number of superconducting vortices is created in the antidot lattice.  At the final temperature below $T_\text{c}$ (which can be of a high-$T_\text{c}$ material) the applied field is switched-off and the vortices remain. 

\section{Magnetic lattice}

In Sec.~\ref{Sec:Poisson} we solve the Poisson equation in order to obtain the magnetic potential generated by an arbitrary array of magnetic monopoles in order to approximate the magnetic field of the superconducting vortex lattice. In Sec~\ref{Sec:perp} we numerically validate the assumption that the perpendicular bias field $\BB_{1}$ is homogeneous close to a sufficiently thin film, as considered in the main text.  In Sec.~\ref{Sec:localaddressing} we analytically (in the London limit) obtain the field generated by a magnetic dipole situated near a superconducting film in all-regions of space.

\subsection{Solution of the Poisson equation} \label{Sec:Poisson}

Let us consider a 2D Bravais lattice of monopoles placed at the lattice points defined with position vectors $\RR$ of the form
\be
\RR=n_{1} \bold{a}_{1} + n_{2} \bold{a}_{2},
\ee
where $\bold{a}_{1(2)}$ are the lattice primitive vectors and $n_{1(2)}$ range through all integer values. The lattice is placed at the plane $z=0$ . The charge distribution is given by
\be
\rho(\rr,z) = 2 \Phi_{0} \delta \left(z \right) \sum_{\RR} \delta \left( \rr- \RR  \right) .
\ee
The factor $2 \Phi_{0}$ appears in connection to the charge associated to a  vortex, recall \eqcite{B:M}. Note that in the main text we consider the monopoles to be placed at $z=-d_\text{m}$, here we consider them to be at $z=0$ to ease the notation. We will just make a change of coordinates at the end of the derivation. The vector $\rr$ is two-dimensional and is defined in the plane $z=0$. The Dirac delta is also two-dimensional.

If a function $f(\rr)$ has the periodicity of a Bravais lattice, \ie $f(\rr + \RR)=f(\rr)$, it can be expressed as
\be
f( \rr) =\sum_{\KK} f_{\KK} e^{\im \KK \cdot \rr}
\ee
where the sum is over all reciprocal lattice vectors $\KK$. The Fourier coefficients $f_{\KK}$ are given by
\be
f_{\KK} = \frac{1}{a^{2}}  \int_{C} d \rr e^{- \im \KK \cdot \rr} f(\rr)
\ee
where the integral is over any direct lattice primitive cell $C$, and $a^{2}$ is the area of the primitive cell. Thus,
\be \label{eq:charge2}
\rho(\rr,z) = 2 B_{0} \delta \left(z \right) \sum_{\KK} e^{- \im \KK \cdot \rr}
\ee
where we have defined $B_{0}= \Phi_{0}/a^{2}$, as in the main text.

Let us solve the Poisson equation
\be
\nabla^{2} \phi(\rr,z)= - \rho(\rr,z)
\ee
for the charge distribution given in \eqcite{eq:charge2}. We will first obtain the potential at the region outside the charge distribution (namely outside the plane at $z=0$) by solving the Laplace equation
\be
\nabla^{2} \phi(\rr,z) =0.
\ee
We will obtain two solutions, one valid at $z>0$ and the other at $z<0$, namely
\be
       \phi(\rr,z)= \left\{ \begin{array}{ll}
               \phi^{+}(\rr,z), & z>0 \\
               \phi^{-}(\rr,z), & z<0.\end{array} \right.
\ee
Since the potential has to be continuum at $z=0$, one requires that
\be
\lim_{\epsilon \rightarrow 0}  \left[ \phi^{+}(\rr,\epsilon) - \phi^{-}(\rr,-\epsilon) \right] = 0.
\ee
The $z$-derivative of the potential is not continuous and yields a boundary condition. The discontinuity can be obtained by integrating the Poisson equation in the nearby of $z=0$, namely
\be \label{eq:poisson}
\begin{split}
\lim_{\epsilon \rightarrow 0} \int_{-\epsilon}^{\epsilon}\nabla^{2} \phi(\rr,z) dz&= - \lim_{\epsilon \rightarrow 0} \int_{-\epsilon}^{\epsilon} \rho(\rr,z) d z \\
&= - 2 B_{0}  \sum_{\KK} e^{- \im \KK \cdot \rr}.
\end{split}
\ee
Now by using that $\partial^{2} \phi(\rr,z)/\partial x^{2}$ and $\partial^{2} \phi(\rr,z)/\partial y^{2}$ are continuous at $z=0$, one is left with the following condition
\be
\begin{split}
&\lim_{\epsilon \rightarrow 0}  \left[ \frac{\partial}{\partial z} \phi^{+}(\rr,z) \large |_{z=\epsilon} - \frac{\partial}{\partial z} \phi^{-}(\rr,z) \large |_{z=-\epsilon}  \right] \\
&= - 2 B_{0} \sum_{\KK} e^{- \im \KK \cdot \rr}.
\end{split}
\ee
In the long-distance limit, the potential has to fulfill 
\be
- \lim_{z \rightarrow \pm \infty} \frac{ \partial \phi(\rr,z)}{\partial z} = \pm B_{0}.
\ee
This can be shown by computing the magnetic field at distance $z \gg a$. In the long distance, the magnetic field has only $z$ component and can be computed by integrating the field created by the infinite amount of monopoles, namely
\be
\begin{split}
&B_{z}(\rr,z \gg a) = \frac{B_{0}}{2 \pi}\sum_{\RR} \frac{  z}{\left[ \left| \rr-\RR \right|^{2}  +  z^{2} \right]^{3/2}} =B_{0}.
\end{split}
\ee

Let us now make the following ansatz for the magnetic potential at $z>0$
\be \label{eq:ansatz1}
\phi^{+}(\rr,z)= \sum_{\KK} e^{\im \KK \cdot \rr}f^{+}_{\KK}(z).
\ee
By introducing \eqcite{eq:ansatz1} into the Laplace equation, one obtains the following differential equation
\be \label{eq:F}
 \partial^{2}_{z} f^{+}_{\KK} (z) =  \frac{f^{+}_{\KK} (z)}{|\KK|^{2}}.
\ee
The general solution of \eqcite{eq:F} is given by
\be
\begin{split}
f^{+}_{\KK}(z) =& A^{+}_{\KK} e^{-z |\KK|} + B^{+}_{\KK} e^{z |\KK|} + C^{+}_{\KK} z. 
\end{split}
\ee
Recalling that 
\be
- \lim_{z \rightarrow \infty} \frac{ \partial \phi(\rr,z)^{+}}{\partial z} = B_{0},
\ee
one obtains the following conditions
\be
\begin{array}{ll}
               B^{+}_{\KK}=0 &  \forall \; \KK \\
               C^+_{\bold{0}}=-B_{0} & \\
               C^+_{\KK}=0 & \forall \; \KK \neq 0  .
\end{array} 
\ee
Thus, the potential can be written as
\be
\begin{split}
 \phi^{+}(\rr,z)= - B_{0}z + \sum_\KK A^{+}_{\KK} e^{\im \KK \cdot \rr} e^{-z |\KK|} 
\end{split}
\ee
The case $z<0$ is analogous and yields 
\be
\begin{split}
 \phi^{-}(\rr,z)=  B_{0} z + \sum_{\KK} A^{-}_{\KK} e^{\im \KK \cdot \rr} e^{z |\KK|} 
\end{split}
\ee
The continuity conditions yields
\be
A^{+}_{\KK} = A^{-}_{\KK} \equiv A_{\KK}
\ee
The discontinuity condition of the $z$-derivative of the potential yields
\be
A_{\KK} =  \frac{B_{0}}{|\KK|} \hspace{1em} \KK \neq \bold{0}
\ee
Putting everything together, the potential is given by
\be
\begin{split}
 \phi^{\pm}(\rr,z) = \mp B_{0} z +   B_{0}\sum_{\KK \neq \bold{0}} \frac{1}{|\KK|}  e^{\im \KK \cdot \rr} e^{ \mp z |\KK| }.
\end{split}
\ee
Since inside the sum over $\KK$ all the terms but the exponential depend on $|\KK|$ and the sum does not include the zero, we can just rewrite it as 
\be
\begin{split}
 \phi^{\pm}(\rr,z) = \mp B_{0} z +   B_{0} \sum_{\KK \neq \bold{0}} \frac{1}{|\KK|}  \cos ( \KK \cdot \rr ) e^{ \mp z |\KK| }.
\end{split}
\ee
From this we can obtain the components of the magnetic potential as given in the main text (by recalling to make a change of coordinates to displace the origin of $z$ from  $z=0$ to $z=-d_\text{m}$.

\subsection{Perpendicular field in a superconducting disk of finite thickness and radius} \label{Sec:perp}

To evaluate how much magnetic field passes through a superconductor in an
externally applied field we will use the following approach. Consider a
cylindrical superconductor of length $L_\text{s}$ and radius $R_\text{s}$ and a uniform applied
field $B_a$ directed along the axis of the cylinder. We set $z=0$ in one of the
planar faces (bottom face) of the cylinder. The
superconductor is modeled within the London approximation which means that the
supercurrents induced in the superconductor, together with the magnetic
interaction, also have kinetic energy~\cite{DeGennesBook}. 

To calculate these currents and, from them, the total magnetic field, we use a
numerical procedure based on minimization of energy 
that simulates the London approximation for the superconductor.

Consider the superconductor to be discretized in $N$ rings of current with
squared cross section of surface $\mathcal{A}$ filling up the superconducting
cylinder. The London energy $G$ can be written as \cite{Chen08}
\begin{eqnarray}
    G&=&  \frac{B_a}{2} \sum_{i=1}^N I_i \rho_i^2 2\pi \nonumber \\ &+&
\sum_{i=1}^N L_i I_i^2 + 
\frac{1}{2}\sum_{i=1}^N\sum_{j=1,j\neq i}^N M_{ij} I_i I_j \nonumber \\ &+&
\frac{\mu_0}{2} \lambda^2 \sum_{i=1}^N I_i^2 \rho_i \frac{2\pi}{\mathcal{A}}
\label{energy}
\end{eqnarray}
where $M_{ij}$ are the
mutual inductance coefficients between the circuits $i$
and $j$, $L_i$ is the self-inductance coefficient of the circuit $i$, $\lambda$
is the London penetration depth and $\rho_i$ are the radial (in cylindrical
coordinates) coordinates of the circuit $i$ and $I_i$ the intensity passing
trough it. Analytical expressions for $M_{ij}$ and $L_i$ can be found in
\cite{LandauBook,JacksonBook}. The first term in Eq.~(\ref{energy}) is the
energy coming from the interaction of the current with the applied field, the
second and third come from the interaction of the currents with themselves and
the last term is the kinetic energy. The minimization of this energy, for a
given applied field yields the current distribution inside the superconductor.
The minimization procedure is performed by finding among all the circuits the
one in which a given small variation of its current minimizes the most
the energy, and repeating the process until no variation can further minimize
the energy \cite{Sanchez01,Chen08}.

\begin{figure}[t]
\centering
\includegraphics[width=\columnwidth]{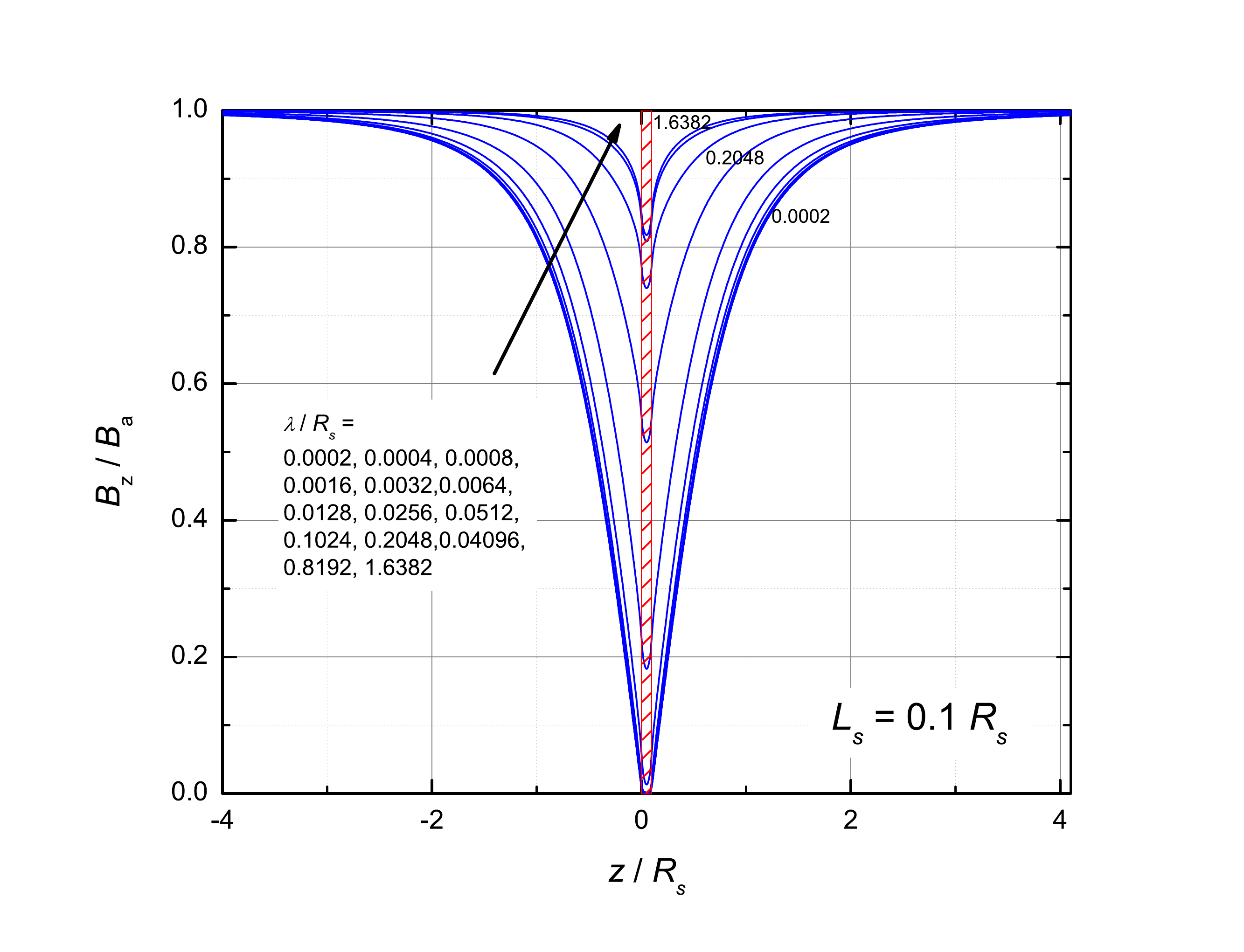}
\caption{Magnetic field (axial component) normalized to the applied field along
the axis of a finite cylinder with $L_\text{s}=0.1R_\text{s}$ for different values of $\lambda$.
The central dashed region represent the space occupied by the superconductor.}
\label{fig1cyl1}
\end{figure}

\begin{figure}[t]
\centering
\includegraphics[width=\columnwidth]{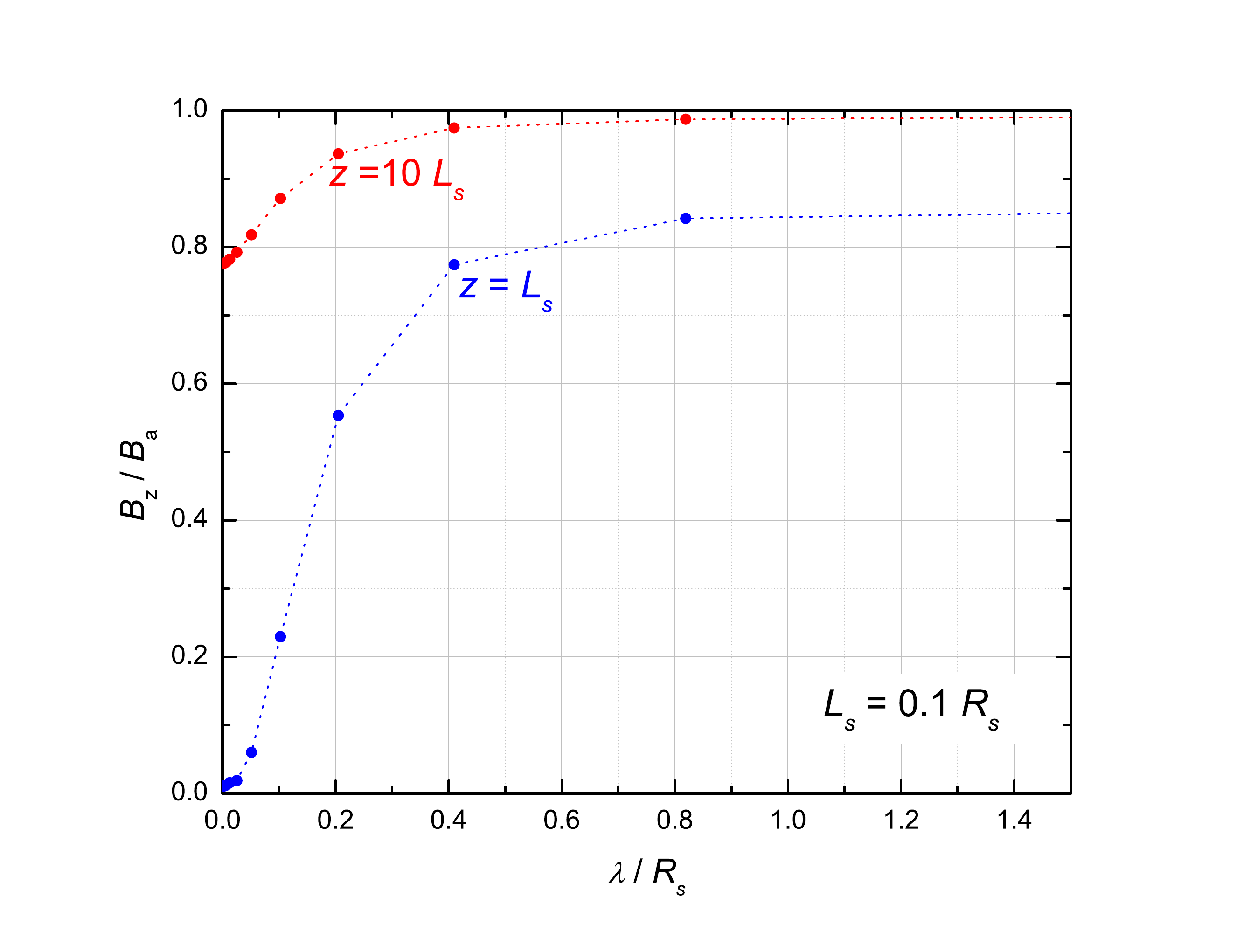}
\caption{Magnetic field (axial component) normalized to the applied field on
the axis of the cylinder evaluated at two vertical distances (blue) $z=L_\text{s}$
and (red) $z=10 L_\text{s}$ as a function of $\lambda/R_\text{s}$. }
\label{fig1cyl2}
\end{figure}

From these calculations, we have found the current distribution in the
superconductor, from which the field everywhere in the space can be calculated
using Biot-Savart law. In Fig.~\ref{fig1cyl1} we
show the field calculated along the axis of the cylinder for different values
of $\lambda/R_\text{s}$ for a given cylinder aspect ratio $L_\text{s}/R_\text{s}=0.1$. In particular to see
how much field crosses the superconductor we plot in Fig.~\ref{fig1cyl2} the
field calculated on the axis and just on top of the superconductor ($z=L_\text{s}$) and
at some distance ($z=10L_\text{s}$) from it. We observe that when $\lambda\rightarrow 0$
we get complete exclusion of the field from inside the superconductor whereas as
$\lambda$ increases, field can pass through the superconductor. When $\lambda$
is large enough, the superconductor is almost transparent to the applied field.

\subsection{Local addressing with a magnetic tip} \label{Sec:localaddressing}

\begin{figure}[t]
\centering
\includegraphics[width=\columnwidth]{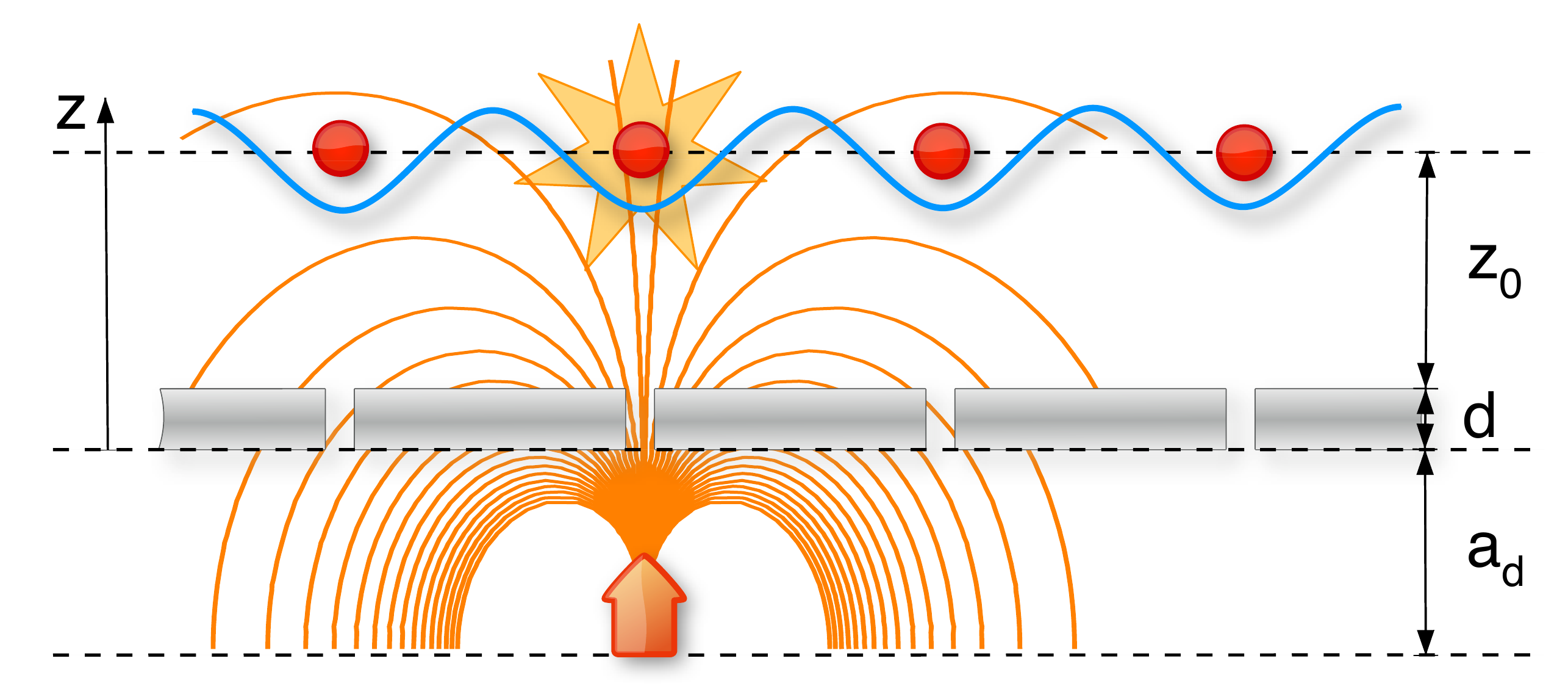}
\caption{Schematic illustration of the local addressing with a magnetic tip. }
\label{fig5b}
\end{figure}

Let us assume now that a magnetic dipole with magnetic moment $\mm_\text{d} =
m_\text{d}(0,0,1)$ is placed below a thin film of thickness $d$ at position
$\rr_\text{d}=(0,0,-a_\text{d})$ (note that the origin of
coordinates is located at the bottom surface of the superconductor, see Fig.~\ref{fig5b}). We
assume that the superconductor does not contain any vortex and is not
necessarily thin. We calculate the magnetic field everywhere in the space
assuming the superconductor to be described by the London equation (without vortices)
\be
{\bf
J}=-\frac{1}{\lambda^2 \mu_0} {\bf A},
\ee
 which together with the Ampere law
(assuming no displacement currents), give us
the equation for the vector potential inside the superconductor (region 2,
$0<z<d$)
\be
\nabla\times (\nabla\times {\bf A}_2) + \frac{1}{\lambda^2} {\bf A}_2 = 0. 
\ee
In region 3 ($z>d$) the magnetic potential vector satisfies
\be
\nabla\times (\nabla\times {\bf A}_3)  = 0.
\ee
In region 1 ($z<0$), for the sake of simplicity, we consider ${\bf A}_1= {\bf
A}_\textrm{dip} + {\bf A'}_1$ so that 
\be
\nabla\times (\nabla\times {\bf A'}_1)  = 0,
\ee
and where ${\bf
A}_\textrm{dip}$ is the magnetic vector potential of a point dipole. In the
present case 
\be
\label{eq:dipole}
\begin{split}
{\bf A}_\textrm{dip}(\rho,z) &= \hat{\bf \phi} \frac{\mu_0 m_\text{d}}{4\pi} \int_0^\infty k
J_1(k\rho) e^{-k(z+a_\text{d})}  \\
 &= \hat{\bf \phi} \frac{\mu_0 m_\text{d}}{4\pi}
\frac{\rho}{[\rho^2+(z+a_\text{d})^2]^{3/2}}.
\end{split}
\ee
Because of the symmetry of the problem, the magnetic vector potential everywhere
has only angular component (denoted as $A_\textrm{dip}, A'_1, A_2$, and $A_3$,
without the $\phi$ subscript) and no angular dependence. All the above
equations can be solved in terms of Bessel functions:
\be
\begin{split}
 A'_1(\rho,z) &=  \frac{\mu_0 m_\text{d}  }{4 \pi} \int_0^\infty \! \! \! P_1(k) J_1(k\rho) e^{k z} \,{\rm d}k,\\
 A_2(\rho,z) &=  \frac{\mu_0 m_\text{d}  }{4 \pi} \\
& \times  \int_0^\infty  \! \! \! J_1(k\rho) [P_{2p}(k) e^{q z/d} + P_{2n}(k) e^{-q z/d}] \,{\rm
d}k,\\
 A_3(\rho,z) &= \frac{\mu_0 m_\text{d}  }{4 \pi}  \int_0^\infty   \! \! \! P_3(k) J_1(k\rho) e^{-k z} \,{\rm d}k,
\end{split}
\ee
where $q \equiv \sqrt{k^2 d^{2}+ d^{2}/\lambda^2}$.  
By setting the boundary conditions (continuity of magnetic vector potential
and
continuity of $z$-component of the magnetic induction field) at $z=0$ and $z=d$
we arrive to a set of equations for the functions $P_1(k), P_{2p}(k),
P_{2n}(k)$, and $P_3(k)$ that can be solved. The solutions are
\be
\begin{split}
 P_1(k) &=  \frac{A(k) k}{d^{2}} (k^2d^{2} - q^2) e^{-a_\text{d} k}(e^{2q }-1),  \\
 P_{2p}(k)&= \frac{2 A(k)  k^2}{d} (q-kd) e^{-a_\text{d}  k}, \\
 P_{2n}(k)&=\frac{2 A(k)  k^2}{d} (q+kd) e^{-a_\text{d}  k+2q }, \\
 P_3(k)&=\frac{4 A(k)  k^2 }{d^{2}} q e^{-a_\text{d}  k+k d+q},
\end{split}
\ee
where
\be
A(k)Ê\simeq d^{2}\left[ \left(k^{2}d^{2}+q^{2}\right)(e^{2q }-1)+2 k dq (e^{2 q }+1) \right]^{-1}
\ee

The field can be now evaluated everywhere in space. In particular, in the main text we compute the induction field at a given position $(0,z_\text{0}+d)$. The field is (only z-component)
\be
\label{eq.bftot}
\begin{split}
& B_{3,z}(0,z_{0}+d)  = \frac{\mu_0 m_\text{d}  }{4 \pi} \int_0^\infty k P_3(k) e^{-k(d+z_{0})} {\rm d}k \\
&= \frac{\mu_0 m_\text{d}  }{4 \pi d^{3}} \int_0^\infty  4  \frac{A(\tilde k/d)}{d^{2}}  \tilde k^3  q e^{-\tilde k (z_{0}+a_\text{d} )/d+q }  {\rm d} \tilde k,
\end{split}
\ee
where in the last equation we have defined $\tilde k=k/d$. It is straightforward to inspect that \eqcite{eq.bftot} divided by \eqcite{eq:dipole} depends only on the dimensionless variables $\lambda/d$ and $(z_{0}+a_\text{d})/d$.
Simplified expressions for the fields can be obtained in the
$\lambda\gg d$ limit since 
\be
P_3(k)=
\frac{k^2\Lambda e^{-a k}}{k \Lambda +1}.
\ee

If the thin film contains a vortex, the total magnetic field is given by the superposition principle, see for instance \cite{Wei96}. In this statement it is implicitly assumed that the vortex is already
present in the superconductor and that the dipole does not produce extra
vortices. To compute the minimum distance above which the dipole will create a vortex, we can use the results given in Sec.~IV in~\cite{Wei96}. For instance, if we approach a dipole to a thin film, it will create a vortex if the distance is smaller than (see Eq.~(4.11) in \cite{Wei96})
\be
a_{1}= \Lambda \sqrt{\frac{\mu_{0} m_\text{d} }{\Phi_{0} \Lambda} \frac{1}{\ln (\Lambda/\xi)}}.
\ee
This equation is valid for $a_{1} \gg \Lambda$. 

\section{Noise and decoherence}

In this section we provide more details on decoherence. In particular, in Sec.~\ref{Sec:jiggling} we derive the spin-flip and motional rates given in the main text. In Sec.~\ref{Sec:damping} we estimate the damping in the harmonic motion of the atom due to the coupling to the ovedamped vortex. In Sec.~\ref{Sec:randomn} we discuss the effect that some randomness in the position of the vortex has in the magnetic potential.

\subsection{Decoherence due to thermal jiggling of vortices} \label{Sec:jiggling}

In the superconducting vortex lattice proposed in the text, it is clear that a source of magnetic field fluctuations will be given by the thermal jiggling of the pinned vortices. In particular the field generated by the vortices is given by
$\BB_\text{V}(\rr,z,t) = \sum_{\RR} \BB_{\RR+\rr_\RR(t)}(\rr,z) $,
where $\rr_\RR(t)$ is the displacement of the vortex pinned at $\RR$.  In a mean-field approach, the vortex position $\rr_\RR$ is customarily determined by the dynamic equation for the balance of forces exerted on a single vortex ~(see \cite{Pompeo08} and reference therein)
\be \label{eq:dynV}
\eta \dot \rr_\RR + \alpha_\text{H} \hat \nn \times \dot \rr_\RR + k_\text{p} \rr_\RR = \FF_\text{ext}  + \FF_\text{T},
\ee
where $\hat \nn$ is the unit vector along the vortex. The first term is the drag force responsible of dissipation of moving vortices, where $\eta$ is the vortex viscosity coefficient, the second describes the perpendicular Hall force, with the Hall coefficient $\alpha_\text{H}$, and the third the pinning force. In a vortex lattice, the repulsive force with the lattice leads to a simple expression (provided $2 \pi \Lambda \gg a$) for the spring constant $k_\text{p}=  \Phi_{0}^{2}/(2 \mu_{0} \Lambda a^{2})$, see~\cite{Brandt09} and SI. On the right hand side, $\FF_\text{ext}$ accounts for the external forces  acting on the vortex, \emph{e.g.}  due to currents in the film, and $\FF_\text{T}$ is an stochastic force due to thermal fluctuations that fulfills $\avg{\FF_\text{T}(t)}=0$ and $\avg{F^{i}_\text{T}(t) F^{j}_\text{T}(t')  } = \delta_{ij} 2 k_\text{B} T \eta \delta (t-t') $, where $k_\text{B}$ is Boltzmann's constant and $T$ the temperature. The vortex mass plays no role in the dynamics since the inertial term is always negligible next to the viscous force~\cite{Golubchik12}. The characteristic microwave frequency $\omega_\text{d}=k_\text{p}/\eta$, the so-called depinning frequency, marks the crossover between elastic motion, dominant at lower frequencies, and purely dissipative motion, arising at higher frequencies~\cite{Pompeo08}. 

In the following we neglect the Hall term and the external forces in  \eqcite{eq:dynV} such that one obtains the following solution
\be
\rr_\RR(t) = e^{- \omega_\text{d}t} \rr_\RR(0) + \frac{1}{\eta}\int_{0}^{t} e^{- \omega_\text{d}(t-\tau)} \bold{F}_\text{T}(\tau) d \tau.
\ee
 In the long-time limit, one gets
\be \label{eq:rr}
\avg{r^{i}_\RR(t) r^{j}_{\RR'}(t+\tau)}= \frac{k_\text{B} T}{\eta \omega_\text{d}} e^{-\omega_\text{d}\tau} \delta_{ij} \delta_{\RR \RR'}.
\ee
This expression will be used in the following to estimate the spin-flip and motional heating rates.

\subsubsection{Spin-flips}

Spin flips in an atom at position $(\RR,z_{0}+d_\text{m})$ occur when there is a fluctuating magnetic field. The rate of spin-flips between levels $i$ and $f$ is given in time-dependent perturbation theory by~\cite{Henkel99}
\be
\Gamma_{i \rightarrow f} = \sum_{\alpha \beta} \frac{\bra{i} \hat \mu_{\alpha} \ket{f} \bra{f} \hat \mu_{\beta} \ket{i}}{\hbar^{2}} S^{\alpha \beta}_{B}(-\omega_{fi}).
\ee
Here $\omega_\text{if}$ is the frequency between the energy levels ``i'' and ``f''. $S^{\alpha \beta}_{B}$ is the magnetic field fluctuation spectrum defined by
\be
S^{\alpha \beta}_{B} (\omega) = 2 \int_{0}^{\infty}\! \! \cos (\omega \tau) \avg{ \delta B^{\alpha}(t+\tau) \delta B^{\beta} (t) }d\tau
\ee
The field at $(\rr,z_{0}+d_\text{m})$ created by the array of vortices is approximated by
\be
\BB_\text{V}(\rr,t)= \sum_{\RR} \BB_{\RR}(\rr,t), 
\ee
where
\be
\BB_{\RR}(\rr,t) = \frac{\Phi_{0}}{2 \pi} \frac{(\rr- \RR- \rr_\RR(t),z_{0}+d_\text{m})}{\left[|\rr- \RR- \rr_\RR(t)|^{2}+(z_{0}+d_\text{m})^{2} \right]^{3/2}}.
\ee
As discussed before, $\rr_\RR (t)$ is the position fluctuations along the $x-y$ plane of the vortex placed at $\RR$. The magnetic field fluctuations can be obtained by
\be
\delta B^{\alpha}(\rr,z,t)=\sum_{\RR} \sum_{i=x,y} b^{\alpha i}_{\RR} r^{i}_\RR(t), 
\ee
where
\be
b^{\alpha i}_{\RR} (\rr) \equiv \left. \frac{\partial B^{\alpha}_{\RR}(\rr)}{\partial r^{i}_\RR} \right |_{\rr_\RR=0 \;\forall \; \RR}.
\ee
The time-averaged magnetic field correlations are thus given by
\be
\begin{split}
\avg{\delta B^{\alpha}(t+\tau) \delta B^{\beta}(t)} &= \sum_{\RR, \RR'} \sum_{i} b^{\alpha i}_{\RR} b^{\beta i}_{\RR'} \avg{r^{i}_\RR(t+\tau) r^{i}_{\RR'}(t)} \\
&=  \frac{ k_\text{B}T}{\eta \omega_\text{d} } e^{-\omega_\text{d} \tau}  \sum_{\RR} \left( \left[ b^{\alpha x}_{\RR} \right]^{2}+ \left[ b^{\alpha y}_{\RR} \right]^{2} \right) \\
&=  \frac{ k_\text{B}T}{\eta \omega_\text{d} } e^{-\omega_\text{d} \tau} \frac{B^{2}_{0}}{a^{2} \Delta^{4}} \frac{3 \pi^{3}}{2}  \delta_{\alpha \beta}(1+\delta_{\alpha z}),
\end{split}
\ee
where we have used \eqcite{eq:rr}. Then, the magnetic field fluctuation spectrum reads
\be
S^{\alpha \beta}_{B}(\omega) =  \frac{2 k_\text{B}T}{k_\text{p}  } \frac{\omega_\text{d}}{\omega^{2}+\omega_\text{d}^{2}}\frac{B^{2}_{0}}{a^{2} \Delta^{4}} \frac{3 \pi^{3}}{2}  \delta_{\alpha \beta}(1+\delta_{\alpha z}), 
\ee
and the spin-flip rate
\be \label{eq:spinflip}
\begin{split}
\Gamma_{i \rightarrow f} (\rr) =& 3 \pi^{3} \frac{ k_\text{B}T}{k_\text{p}  } \frac{\omega_\text{d}}{\omega_{fi}^{2}+\omega_\text{d}^{2}}\frac{B^{2}_{0}}{a^{2} \Delta^{4}}   \times
\\ & \times \sum_{\alpha} \frac{|\bra{i} \mu_{\alpha} \ket{f} |^{2}}{\hbar^{2}} (1+\delta_{\alpha z}).
\end{split}
\ee

\begin{figure}[t]
\begin{center}
\includegraphics[width= 0.9 \linewidth]{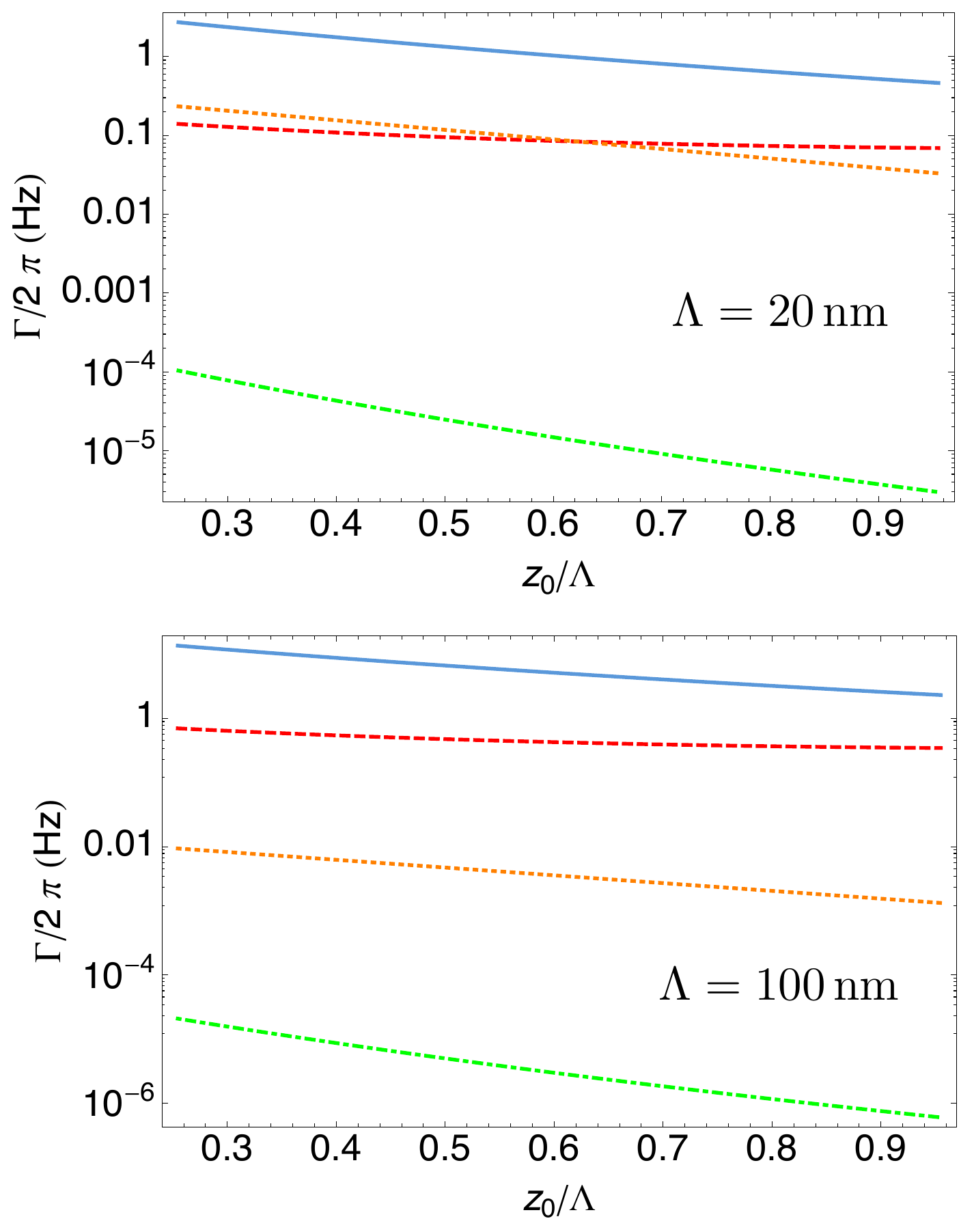}
\caption{The spin-flip rate \eqcite{eq:spinflip} (solid blue line), the motional heating rate \eqcite{eq:heating} (dashed red line), the Majorana losses rate (dotted orange line) $\Gamma_\text{ML}\approx 2\pi \omega_\text{t} \exp[-4 \omega_\text{L}/\omega_\text{t}]$, and the damping rate ~\eqcite{eq:damping} (dotdashed green line) are plotted as a function of $z_{0}/\Lambda$ for $\Lambda=20$ nm and $\Lambda=100$ nm. We considered Lithium and the same parameters given in the caption of Figure 3 in the main text.}
\label{Fig:gamma}
\end{center}
\end{figure}

\subsubsection{Motional Heating}

The magnetic field fluctuations also lead to motional heating due to the fluctuations in the interaction $- x_{0} (\hat b + \hat b^{\dagger}) \bold{n} \cdot \nabla (\hat \mu \cdot \delta \BB(\rr, t))$, where $x_{0}=\sqrt{\hbar/(2 m_\text{a} \omega_\text{t})}$. Using time-dependent perturbation theory, the motional heating rate is given by~\cite{Henkel99}
\be
\Gamma_{0 \rightarrow 1} = \frac{x^{2}_{0}}{\hbar^{2}} \sum_{ij} n_{i}n_{j} S^{ij}_{F}(-\omega_\text{t} ) .
\ee
In this case, the force fluctuation spectrum is given by 
\be
S^{ij}_{F} (\omega) = 2 \int_{0}^{\infty}\! \! \cos (\omega \tau) \avg{ F_{Z}^{i} (t+\tau) F_{Z}^{j}(t) }d\tau,
\ee
where 
\be
\begin{split}
F^{i}_\text{Z}( \rr,z,t) &= \sum_{\alpha}\mu_{\alpha}  \partial_{i} B^{\alpha} (\rr,t) \\
&= \sum_{\alpha,\RR,j}\mu_{\alpha}  \partial_{i} b^{\alpha j}_{\RR} r^{j}_\RR(t) .
\end{split}
\ee
Analogously to the spin-flip calculation, one arrives at
\be
\begin{split}
\avg{F^{i}_\text{Z}(t+\tau) F^{k}_\text{Z}(t) }&= \delta_{ik} \frac{ k_\text{B}T}{\eta \omega_\text{d} } e^{-\omega_\text{d} \tau}   \sum_{\alpha,\RR,j}\mu_{\alpha}^{2}   \left[ \partial_{i} b^{\alpha j}_{\RR}  \right]^{2} \\
&= \delta_{ik} \frac{ k_\text{B}T}{\eta \omega_\text{d} } e^{-\omega_\text{d} \tau}  \pi (2 \pi)^{4}   \frac{B^{2}_{0} }{a^{4} \Delta^{6}} A_{i},
\end{split}
\ee
where
\be
\begin{split}
A_{x}&= \frac{45}{32} \mu^{2}_{x} + \frac{15}{32} \mu^{2}_{y} + \frac{15}{8} \mu^{2}_{z}, \\
A_{y}&=\frac{15}{32}  \mu^{2}_{x} + \frac{45}{32}  \mu^{2}_{y} + \frac{15}{8} \mu^{2}_{z},  \\
A_{z}&=\frac{15}{8} \mu^{2}_{x} + \frac{15}{8} \mu^{2}_{y} + \frac{15}{4} \mu^{2}_{z}.  \\
\end{split}
\ee
Hence, 
\be
S^{ij}_{F}(\omega) = \frac{2 k_\text{B}T}{k_\text{p}  } \frac{\omega_\text{d}}{\omega^{2}+\omega_\text{d}^{2}} \pi (2 \pi)^{4}   \frac{B^{2}_{0} }{a^{4} \Delta^{6}} A_{i}\delta_{ij},
\ee
and 
\be \label{eq:heating}
\Gamma_{0 \rightarrow 1}=(2 \pi)^{5} \frac{  k_\text{B}T}{ k_\text{p}  } \frac{\omega_\text{d}}{\omega_\text{t}^{2}+\omega_\text{d}^{2}}    \frac{x^{2}_{0}B^{2}_{0} }{a^{4} \Delta^{6}}  \sum_{i} \frac{n^{2}_{i}  A_{i}}{\hbar^{2}}.
\ee

\subsection{Atom damping due to coupling to vortex} \label{Sec:damping}

Consider a magnetic dipole oscillating above a thin film superconductor 
containing one vortex. The dipole is at position $(\RR+\rr_\text{a},z_{0}+d_\text{m})$ and has a magnetic moment ${\bf m}_\text{a} = m_\text{a}
\hat z$. The vortex is approximated by a monopole placed at position $(\RR+ \rr_\RR,0)$.
The interaction energy between the dipole and the vortex is given by
\be
\begin{split}
  U_\text{aV} (|\rr_\text{a}-\rr_\RR|&=- {\bf m}_\text{a} \cdot {\bf
B}_\RR (\rr_\text{a},z_{0}+d_\text{m}) \\
&=- \frac{\Phi_0 }{2\pi } \frac{m_\text{a} (z_{0}+d_\text{m})}{\left[(z_{0}+d_\text{m})^{2} + |\rr_\text{a}-\rr_\RR |^{2} \right]^{3/2}}.
\end{split}
\ee
Assuming that $(z_{0}+d_\text{m}) \gg |\rr_\text{a}-\rr_\RR |$, one can approximate
\be \label{eq:UPV}
U_\text{aV}(|\rr_\text{a}-\rr_\RR|) \approx  U_\text{PV}(0) + \frac{1}{2} k_\text{int} |\rr_\text{a}-\rr_\RR|^{2}
\ee
where
\be \label{eq:kint}
k_\text{int} \equiv (2 \pi)^{3} \frac{3 m_\text{a} B_{0}}{a^{2}\Delta^{4}} .
\ee
We used $\Delta = 2 \pi (z_{0}+d_\text{m})/a$ and $B_{0}=\Phi_{0}/a^{2}$, as in the main text.

We assume that the dipole is forced to oscillate along the $x$-axis with an
amplitude $x_{0}=[\hbar/(2 m_\text{a} \omega_\text{t})]^{1/2}$ and period $2\pi /\omega_\text{t}$, namely $\rr_{a}(t)=(x_{0} \sin \omega t, 0)$.  The vortex is assumed to be in a
potential well with lateral spring constant $k_{p}$ and drag coefficient
$\eta$. The magnetic coupling between the dipole and
the vortex induces some oscillating motion to the vortex. We are assuming that
all the movements are done in the $x$-axis. We
would like to calculate the dynamics of the vortex and the power dissipated. As a first approximation, we assume that the oscillation of the dipole is driven, namely, it is not damped due to the vortex dissipation. This yields a faster energy dissipation than the case where the dipole is not driven. Using \eqcite{eq:UPV}, we can write the dynamic equation of the balance forces on the vortex. The
force exerted on the vortex by the dipole is 
\be
F_\text{ext}^{x}=- \frac{\partial U_\text{aV}}{\partial r^{x}_\RR}= - k_\text{int} r^{x}_\RR + f_\text{a}(t),
\ee
where $f_\text{a}(t)\equiv  k_\text{int} x_{0} \sin \omega t$. Taking into account the vortex dynamics neglecting the thermal fluctuations and the Hall term, 
\be
\eta \dot r^{x}_\RR + (k_\text{p}+k_\text{int}) r^{x}_\RR =  f_\text{a}(t).
\ee
In the steady state, the solution of the vortex motion is given by
\be
r^{x}_\RR (t)= X_{0} \sin (\omega_\text{t} t - \varphi),
\ee
where
\be
\tan \varphi = \frac{\eta \omega_\text{t}}{k_\text{p}+k_\text{int}},
\ee
and
\be
X_{0}= \frac{k_\text{int} x_{0}}{\sqrt{(k_\text{p}+k_\text{int})^{2}+\eta^{2} \omega_\text{t}^{2}}}.
\ee
The power dissipated in one period of oscillation is given by
\be
\bar P = \frac{\omega_\text{t}}{2 \pi}\int_{0}^{2 \pi /\omega_\text{t}} dt \dot r^{x}_\RR (t) f_{\text{a}}(t) = \frac{1}{2}\frac{k_\text{int}^2 x_0^2 \eta \omega_\text{t}^2}{(k_\text{p}+k_\text{int})^2+\eta_\text{p}^2\omega_\text{t}^2}.
\ee
 A relevant rate for decoherence is  $\Gamma_\text{d}=1/t^{\star}$, where $t^\star$ is the time at which the energy dissipated is equal to $\hbar \omega$, namely
\be  \label{eq:damping}
\Gamma_\text{d}= \frac {\bar P} {\hbar \omega_\text{t}} = \frac{1}{2 \hbar} \frac{k_\text{int}^2 x_0^2 \eta \omega_\text{t}}  {(k_\text{p}+k_\text{int})^2+\eta_\text{p}^2\omega^2}\approx \frac{x_{0}^{2} \eta \omega_\text{t}} {2\hbar} \frac{k^{2}_\text{int}} {k_\text{p}^{2}}.
\ee
where in the last equation we used that with typical numbers $k_\text{p} \gg k_\text{int}$.

See Fig.~\ref{Fig:gamma} for a comparison of the spin-flip rate, the motional heating rate, and the damping rate. 

\begin{figure}[t]
\begin{center}
\includegraphics[width= 0.7 \linewidth]{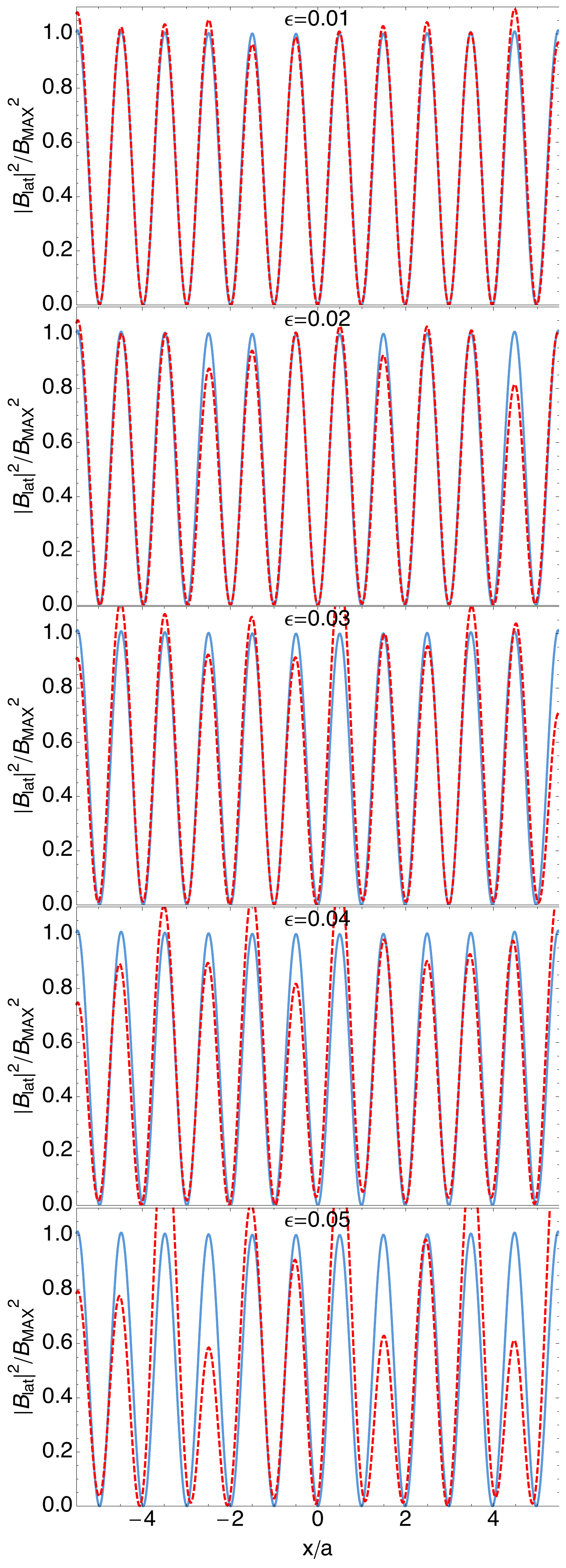}
\caption{$|\BB_\text{lat}|^{2}$ is plotted for $\Delta=2 \pi(z_{0}+d_\text{m})/a=\pi$ at $z=z_{0}$ and $y=0$ as a function of $x/a$ considering a random positioning of vortices with errors given by $\epsilon=0.01,0.02,0.03,0.04,0.05$ (dashed red line). The perfect case $\epsilon=0$ is plotted with a solid blue line. $|\BB_\text{lat}|^{2}$ is plotted in units of $B_\text{max}\equiv |\BB^{2}_\text{lat}(x=a/2,0,z_{0})|^{2}$ (in the case of $\epsilon=0$). We have used a square lattice of $201 \times 201$ vortices. }
\label{Fig:error}
\end{center}
\end{figure}

\subsection{Randomness in vortex position} \label{Sec:randomn}

In the construction of the vortex lattice, we considered that the antidots, and therefore also the vortices, will be positioned at the Bravais lattice points $\RR$. In a real experiment, there will be some error in this positioning, namely, the vortex at position $\RR$ will be sit in reality at $\RR+\rr_\RR$, where $r^{i}_\RR \in a (-\epsilon,\epsilon)$ is a random number. Hence, $\epsilon$ parametrizes the error in positioning the vortices. This error will have an effect in the magnetic field created by the superconducting vortex lattice
$\BB_\text{V}(\rr)= \sum_{\RR} \BB_{\RR+\rr_\RR}(\rr)$, 
where
\be
\BB_{\RR+\rr_\RR}(\rr) = \frac{\Phi_{0}}{2 \pi} \frac{(\rr- \RR- \rr_\RR,z_{0}+d_\text{m})}{\left[|\rr- \RR- \rr_\RR(t)|^{2}+(z_{0}+d_\text{m})^{2} \right]^{3/2}}.
\ee
In Fig.~\ref{Fig:error} we plot $|\BB_\text{lat}|^{2}$ (recall that $\BB_\text{lat}=\BB_\text{V}+\BB_{1}$) for $\Delta=2 \pi(z_{0}+d_\text{m})/a=\pi$ at $z=z_{0}$ and $y=0$ as a function of $x/a$ for $\epsilon=0.01,0.02,0.03,0.04,0.05$ in a square lattice of $201 \times 201$ vortices. As can be seen in Fig.~\ref{Fig:error}, the depth of the lattice sites starts to fluctuate significantly for $\epsilon \gtrsim 0.02$. While this lattice fluctuations could be used to study quantum many-body physics in disordered lattice potentials, it might pose an experimental challenge when regular lattices are required. As mentioned in the main text, for the latter purpose a triangular lattice naturally formed in a thin film without pinning might be advantageous.

\end{document}